\documentclass[10pt,twocolumn,a4paper]{article}
\usepackage[a4paper,top=1in,bottom=1in,left=0.8in,right=0.8in,
  columnsep=24.45pt]{geometry} %

\usepackage[nocompress]{cite}
\ifdefined\XeTeXversion
  \usepackage{fontspec}
\else
  \usepackage[T1]{fontenc}
  \usepackage[utf8]{inputenc}
\fi
\usepackage{subcaption}
\usepackage{amsfonts}
\usepackage{xcolor}
\usepackage{color}
\usepackage[normalem]{ulem}
\usepackage{stackrel}
\usepackage[linesnumbered,ruled,vlined]{algorithm2e}
\usepackage{amsthm,amsmath}
\usepackage{amssymb}
\usepackage{mdframed}
\newmdenv[
  leftmargin=0.7em,     %
  rightmargin=0.7em,    %
  linewidth=1pt,      %
  innertopmargin=2pt,  %
  innerbottommargin=2pt, %
  innerleftmargin=2pt, %
  innerrightmargin=2pt %
]{boxedquote}
  
\usepackage{thmtools}

\usepackage{xspace}
\usepackage{xifthen}

\let\sigproof\proof\let\proof\relax
\let\sigendproof\endproof\let\endproof\relax
\let\proof\sigproof
\let\endproof\sigendproof

\usepackage{tikz}
\usetikzlibrary{positioning}
\usetikzlibrary{decorations.pathreplacing,calligraphy}
\usetikzlibrary{positioning}
\usetikzlibrary{arrows}
\usetikzlibrary{calc}

\usepackage{graphicx}

\usepackage{booktabs}

\usepackage{tabularx}

\usepackage{enumerate}

\usepackage{url}
\usepackage{etoolbox}
\appto\UrlBreaks{\do\-}

\usepackage{multirow,array}
\newcolumntype{P}[1]{>{\centering\arraybackslash}p{#1}}
\usepackage{tabularray}

\usepackage{hyperref}
\hypersetup{
	colorlinks=true,
	citecolor=blue,
}

\usepackage{cleveref}
\crefname{figure}{Figure}{Figures}
\Crefname{figure}{Figure}{Figures}
\crefformat{section}{\S#2#1#3}
\crefrangeformat{section}{\SS#3#1#4-#5#2#6}
\Crefformat{section}{\S#2#1#3}
\Crefrangeformat{section}{\SS#3#1#4-#5#2#6}
\crefname{thm}{Theorem}{Theorems}
\Crefname{thm}{Theorem}{Theorems}
\crefname{lemma}{Lemma}{Lemmas}
\Crefname{lemma}{Lemma}{Lemmas}
\crefname{assm}{Assumption}{Assumptions}
\Crefname{assm}{Assumption}{Assumptions}
\crefname{cor}{Corollary}{Corollaries}
\Crefname{cor}{Corollary}{Corollaries}
\crefname{defn}{Definition}{Definitions.}
\Crefname{defn}{Definition.}{Definitions.}

\theoremstyle{definition}
\newtheorem{thm}{Theorem}
\newtheorem{lemma}[thm]{Lemma}
\newtheorem{cor}[thm]{Corollary}
\newtheorem{defn}[thm]{Definition}

\newtheorem{manualtheoreminner}{Theorem}
\newenvironment{manualtheorem}[1]{%
  \IfBlankTF{#1}
    {\renewcommand{\themanualtheoreminner}{\unskip}}
    {\renewcommand\themanualtheoreminner{#1}}%
  \manualtheoreminner
}{\endmanualtheoreminner}

\newtheorem{manualcorollaryinner}{Corollary}
\newenvironment{manualcorollary}[1]{%
  \IfBlankTF{#1}
    {\renewcommand{\themanualcorollaryinner}{\unskip}}
    {\renewcommand\themanualcorollaryinner{#1}}%
  \manualcorollaryinner
}{\endmanualcorollaryinner}

\newcommand{\conj}{\mathrel{\wedge}}
\newcommand{\disj}{\mathrel{\vee}}
\newcommand{\defeq}{\mathrel{\triangleq}}

\newcommand{\ndepth}{\ensuremath{\ell}}
\newcommand{\nwidth}{\ensuremath{w}}
\newcommand{\nrouters}{\nwidth}
\newcommand{\sender}{\ensuremath{\mathfrak{s}}}
\newcommand{\senders}{\ensuremath{\mathsf{S}}}
\newcommand{\receiver}{\ensuremath{\mathfrak{r}}}
\newcommand{\receivers}{\ensuremath{\mathsf{R}}}

\newcommand{\router}{\ensuremath{\mathfrak{t}}}
\newcommand{\routers}{\ensuremath{\mathfrak{T}}}
\newcommand{\irouter}[1]{\ensuremath{\router_{#1}}}
\newcommand{\mrouter}[2]{\ensuremath{\router_{#1,#2}}}
\newcommand{\msender}[1]{\ensuremath{\sender_{#1}}}

\newcommand{\rauth}{\ensuremath{\mathfrak{R}}}

\newcommand{\rind}{\ensuremath{h}}

\newcommand{\lind}{\ensuremath{i}}
\newcommand{\lindB}{\ensuremath{j}}

\newcommand{\mind}{\ensuremath{n}}
\newcommand{\mindB}{\ensuremath{m}}
\newcommand{\mindC}{\ensuremath{a}}
\newcommand{\mindD}{\ensuremath{b}}

\newcommand{\msg}{\ensuremath{\mathfrak{m}}}

\newcommand{\nmsgs}{\ensuremath{N}}

\newcommand{\pcompromise}{\ensuremath{\alpha}}

\newcommand{\prd}[2]{\ensuremath{\mbox{Pr}_{#1}\left[#2\right]}}

\newcommand{\expec}[1]{\ensuremath{\mathbb{E}\left[#1 \right]}}
\newcommand{\expecd}[2]{\ensuremath{\mathbb{E}_{#1}\left[#2 \right]}}

\newcommand{\covarn}[1]{\ensuremath{\mathsf{Cov}\left[#1\right]}}
\newcommand{\varnd}[2]{\ensuremath{\mathsf{Var}_{#1}\left[#2\right]}}
\newcommand{\obs}{\ensuremath{O}}

\newcommand{\obss}{\ensuremath{o}}

\newcommand{\pcond}[2]{\ensuremath{\left. #1 \;\middle|\; #2\right.}}

\newcommand{\postp}{\ensuremath{\mathfrak{p}}}
\newcommand{\postpr}[2]{\postp_{#1,#2}}
\newcommand{\postpB}{\ensuremath{\mathfrak{q}}}

\newcommand{\dpep}{\varepsilon}
\newcommand{\dpde}{\delta}

\newcommand{\sample}{\mathrel{\overset{{\scriptscriptstyle\$}}{\leftarrow}}}

\newcommand{\nrmsgs}[2]{\ensuremath{\nmsgs_{#1,#2}}}

\newcommand{\incidents}[2]{\ensuremath{I_{#1,#2}}}

\newcommand{\exigents}[2]{\ensuremath{O_{#1,#2}}}

\newcommand{\tgtmsg}[1]{\ensuremath{\overline{\msg}_{#1}}}

\newcommand{\idealarr}[1]{\ensuremath{G}_{#1}}
\newcommand{\idealarrs}[1]{\ensuremath{S}_{#1}}

\newcommand{\distbinom}{\ensuremath{\mathrm{Bin}}}
\newcommand{\mdistbinom}[2]{\ensuremath{\distbinom\left(#1; #2\right)}}

\newcommand{\randomness}{\ensuremath{\mathfrak{x}}}

\newcommand{\permglob}{\ensuremath{\randomness^{G}}}
\newcommand{\mpermglob}[1]{\ensuremath{\permglob_{#1}}}

\newcommand{\cvar}{\ensuremath{\mathcal{V}}}
\newcommand{\mcvar}[1]{\ensuremath{\cvar_{#1}}}

\newcommand{\cmean}{\ensuremath{\mu}}
\newcommand{\mrcmean}[2]{\ensuremath{\cmean_{#1,#2}}}

\newcommand{\meandiff}{\ensuremath{\mathfrak{e}}}
\newcommand{\mrmeandiff}[2]{\ensuremath{\meandiff_{#1,#2}}}

\newcommand{\partinf}{\ensuremath{\mathcal{F}}}
\newcommand{\roundinf}{\ensuremath{\mathcal{H}}}

\newcommand{\contfact}[2]{\ensuremath{\gamma_{#1,#2}}}

\newcommand{\samerouter}[2]{\ensuremath{[\![#2]\!]_{#1}}}
\newcommand{\sentto}[2]{\ensuremath{\router\router_{#1}(#2)}}
\newcommand{\indfn}[1]{\ensuremath{\mathbf{1}_{#1}}}

\newcommand{\arsinh}{\operatorname{arsinh}}

\renewcommand{\paragraph}[1]{\medskip\noindent\textbf{#1.}}

\begin{document}

\title{Provable Mixing Properties of Parallel Mixnets Revisited}
\author{%
  Pierfrancesco Ingo, Peter Druschel and Deepak Garg\\
  \{pieringo,dg,druschel\}@mpi-sws.org\\
  \mbox{}\\
  Max Planck Institute for Software Systems
  }
\date{}
	
\maketitle
\begin{abstract}
A mix-network (\emph{mixnet}) is a routing network that breaks
source-to-destination message trackability by indistinguishably
``mixing'' the routes of concurrently transmitted messages, thus
providing anonymity to senders, receivers and sender-receiver pairs.
Given their potential use in high-stakes scenarios like protecting
whistleblowers, one would like to prove formally that sender anonymity
holds in a mixnet, even against strong adversaries that can compromise
subsets of mixnet servers. However, existing work on anonymity in
mixnets is either limited to linear mixnets (which do not scale
horizontally), relies on special provisions such as additional noise
or chaff messages, or is empirical -- typically based on entropy,
which does not provide strong guarantees  if the adversary has
auxiliary information.

In this paper, we present a new analysis of \emph{parallel} mixnets
which scale horizontally, showing that these networks can provide
strong indistinguishability for messages even without additional noise
messages. Our analysis works by bounding the \emph{statistical
  variance of the adversary's posterior likelihood} of where a message
of interest is at a given point in time. Using concentration bounds on
this distribution, we show that parallel mixnets satisfy
$(\varepsilon, \delta)$-differential indistinguishability for messages routed
through the mixnet (similar to $(\varepsilon, \delta)$-differential privacy)
with very small failure probabilities $\delta$ and small privacy
parameters $\varepsilon$ under reasonable operating conditions. Our model
also enables an examination of the quantitative trade-off between
anonymity, latency and per-router bandwidth in parallel mixnets.
\end{abstract}

\section{Introduction}

Mix-networks or \emph{mixnets} are a fundamental building block of
anonymous communication~\cite{diaz03towards,
  diaz24mixmatch,diaz25shape,Donghang23rpm,Daichong24anonsurvey,rahimi25mocha,rahimi25parsan,
  kate25tightrope,
  lazar18karaoke,dingledine04tor,guirat22mixnetopt,loopix17mixnet,vuvuzela15mixnet,nym21whitepaper,mixim21mixnetsim,
  danezis03towards, chaum17cmix,diaz10impact, barman22groove,
  gibbs15riposte, kwon17atom, kwon20xrd, tyagi16stadium}. Mixnets
provide anonymity by routing all messages through a set of servers;
each server indistinguishably mixes all messages it receives within a
round, a fixed duration of time. The ideal mixnet consists of a single
server that collects messages from senders in each round, shuffles
them, and forwards them to recipients with reencryption. Assuming that
$\nmsgs$ senders send messages in a round, the adversary's
\emph{posterior} distribution over possible senders of a given message
exiting the network is uniform with probability mass $1/\nmsgs$, i.e.,
the ideal mixnet provides perfect anonymity. However, the ideal mixnet
has two shortcomings: (1) it has a single point of failure --- its
only server --- and (2) its throughput is limited to that of one
server.

These shortcomings are addressed by a simple generalization called a
\emph{parallel mixnet}, which consists of $\nwidth$ servers that can
communicate with each other pairwise. To send a message, the sender
selects a random sequence of $\ndepth$ servers from this set and sends
its message to the first server in this sequence in the next available
round.\footnote{The discussion here also applies to a parallel mixnet
  that is structurally separated into $\ndepth$ layers, each with
  $\nwidth$ distinct servers. Our analysis assumes this layered
  structure (\cref{sec:threat-model}).} At the end of each round, every server
uniformly mixes the messages it has received in that round, and
forwards each message to the next server on the message's chosen path,
after re-encryption (the last server on a message's path delivers the
message to the intended recipient). A parallel mixnet scales
horizontally with the number of servers because the average
steady-state load on a server is $\nmsgs/\nwidth$, which reduces with
$\nwidth$, and a parallel mixnet's anonymity degrades gracefully with
the fraction of servers the adversary can compromise. At the same
time, because each message mixes indistinguishably with only a subset
of other messages in each round (exactly the subset that share the
router in that round), the anonymity provided by a parallel mixnet is
probabilistic and not as high as that provided by the ideal mixnet.

We are interested in the rigorous, provable analysis of properties of
parallel mixnets and, specifically, the following two questions:

\medskip
\noindent
\textbf{RQ1.} What \emph{provable}, quantitative anonymity
property(ies), expressed in terms of $\nmsgs$, $\nwidth$ and
$\ndepth$, does a parallel mixnet offer?

\medskip
\noindent
\textbf{RQ2.} What is the quantitative trade-off between the anonymity
properties and performance (bandwidth required, message latency) of a
parallel mixnet in terms of the same parameters?

Prior work in this space has made partial progress towards answering
these two questions, but never provided a definitive answer,
particularly when it comes to \emph{provable lower bounds on strong
  anonymity}. In the following, we outline closely related prior work
in this space and then describe what we do to fill the gap.

\paragraph{Prior work}
A pioneering, foundational baseline is a quantitative tradeoff
between strong anonymity, low bandwidth and low latency due to Das
\emph{et al.}~\cite{DBLP:conf/sp/Das0MK18}, who establish an upper
bound (not a lower bound) on the anonymity that a mixnet can provide
for any fixed values of bandwidth and latency. This bound is quite
loose as it is derived from general arguments, not a specific analysis
of parallel mixnets, and it is orders of magnitude away from
empirically estimated anonymity (see~\cite[Table
2]{kate25tightrope}). At the other extreme, there are principled
methods to \emph{estimate} the degree of anonymity afforded by a
parallel mixnet very precisely using empirical simulations, even with
confidence intervals~\cite{kate25tightrope,diaz03towards}, but these
methods do not yield provable bounds.

Other lines of work have analyzed \emph{specific} parallel mixnet
designs that add more structure to the vanilla parallel mixnet
described above. Karaoke~\cite{lazar18karaoke} and systems related to
it such as Yodel~\cite{yodel}, Groove~\cite{barman22groove} and
Stadium~\cite{tyagi16stadium} are highly scalable parallel mixnets
that rely on additional noise (chaff) messages between pairs of
servers to create specific ``gadgets'' (patterns of message flows)
with high probability. Analyzing the probability with which such
gadgets exist and relying on a relational argument, Karaoke proves
that any two sent messages become indistinguishable to an adversary
with very high probability after a few rounds of mixing. However, the
argument does not work without the chaff messages, which number in
$O((\nwidth \cdot \ndepth)^2)$, and the proof is unsound as argued
with a counterexample in later
work~\cite{DBLP:conf/csfw/DasMMK24}. The same later work proves strong
anonymity for a different mixnet design, but that design also works
quite differently from the vanilla parallel mixnet.

Thus far, there is no provable lower bound on strong anonymity
properties of the vanilla parallel mixnet. Our work fills this gap.

\paragraph{Our contributions}
We present a new stochastic analysis of the routing within the
barebones parallel mixnet to establish a provable lower bound on the
strong anonymity it provides as a function of the operational
parameters $\nwidth$, $\ndepth$, $\nmsgs$ and the probability that any
given server is compromised by the adversary (denoted $\pcompromise$
later). Our bound does not assume any chaff messages or specialized
routing strategy. We further show that for practical values of the
operational parameters, our bound is within the confidence bands of
the empirical estimates of anonymity from prior
work~\cite{kate25tightrope,kate13anoa}.

We work with a strong anonymity property inspired by differential
privacy that is also used by prior work on parallel
mixnets~\cite{kate25tightrope}.  The property, which we call
\emph{$(\dpep, \dpde)$-message indistinguishability}, works as
follows. Suppose that a rational adversary wants to determine where a
specific entering message exits the mixnet in a specific run of the
mixnet. Based on its observation of message routing during the run,
the adversary can assign \emph{posterior odds} to each message in the
network, representing its a posteriori belief that this message is the
message it is tracking. \cref{fig:forward-probs-abstract} illustrates
how the adversary calculates these odds after observing the full
routing in a run of the system (a posteriori): At each routing step on
a server, the adversary adds all the incoming odds on the server and
divides them evenly among the outgoing messages.

The $(\dpep, \dpde)$-message indistinguishability property states
that, except with a very small probability, the posterior odds on any
two messages after the last layer of routing are very close to each
other. In security terms: Even if the adversary is told by an oracle
that the message of interest is one of only two exiting messages, the
adversary's posterior odds on the two messages, say, $p_1$ and $p_2$,
are nearly equal. Specifically, except with probability at
most $\dpde$ over runs, $p_1 \leq e^\dpep \cdot p_2$
for very small $\dpep$ and $\dpde$ (ideally, of the orders of $10^0$
or less and $10^{-6}$ or less, respectively). This is a very strong
anonymity property as it establishes message untraceability even when
the adversary has a huge amount of auxiliary information that rules
out $\nmsgs-2$ of $\nmsgs$ exiting messages. To establish this
property, we have to find $\dpep$ and $\dpde$ in terms of $\nwidth$,
$\ndepth$, $\nmsgs$ and $\pcompromise$, and prove that they indeed
satisfy $p_1 \leq e^\dpep \cdot p_2$ with
probability at least $1 - \dpde$ for all $p_1$ and $p_2$
in the last layer.

Our \emph{key insight} is that, through a careful but elementary
stochastic analysis of routing in a parallel mixnet, it is possible to
\emph{express the variance of the distribution of the calculated
  posterior odds} after the $k$th layer of routing as a closed-form
expression over $\nmsgs$, $\nwidth$, $\ndepth$ and
$\pcompromise$. Given this formula for the variance, we can use a
concentration bound like Chebyshev's inequality to derive a tail
probability bound on each posterior odds value at $k = \ndepth$ and, from
that, an upper bound on the ratio $p_i/p_j$. The tail probability is
our required $\dpde$, and the upper bound on the ratio is our
$\dpep$. Together, they constitute a minimum guaranteed degree of
anonymity afforded by the parallel mixnet.

Having expressed $\dpep$ and $\dpde$ in terms of our operational
parameters, we can also easily explore the anonymity-bandwidth-latency
tradeoff of the mixnet as bandwidth per router is directly
proportional to $\nmsgs/\nwidth$ and message latency is directly
proportional to $\ndepth$.

In technical terms, our stochastic analysis extends that of Goodrich
and Mitzenmacher (GM)~\cite{goodrich12cardspub}, who show that the
\emph{sum} of the variances of the posterior odds on messages exiting
the mixnet matching a given input message falls exponentially in the
number of mixing rounds, $\ndepth$. Hence, all odds converge to their
mean $1/\nwidth$. However, the GM analysis is limited to an idealized
case where the messages are \emph{assumed} to be evenly spread across
the servers in each forwarding round, which is untrue in general. In
an unpublished extension~\cite[Section
4.3]{goodrich2012anonymouscardshufflingapplications}, GM extend their
analysis to alleviate this restriction, but they leave the expression
for the reduction factor in a combinatorial form with nested sums and
do not examine its properties or consequences any further, precluding
immediate use of the extension. Our analysis goes significantly
further: We observe how the sum of variances leads to an exact,
closed-form expression for the variance of every posterior odds value and how
that leads to $(\dpep, \dpde)$-message indistinguishability.

\paragraph{Summary of contributions}
We summarize the contributions of our work.
\begin{itemize}
\item A provable, quantitative lower bound on strong anonymity
  provided by a vanilla parallel mixnet.
\item A new stochastic analysis of parallel mixnets to rigorously
  prove the bound.
\item An exploration of the practical implications, confirming that
  parallel mixnets provide excellent anonymity in practice, with good
  performance, but now with formal proof of anonymity.
\item A comparison of our provable anonymity bounds to those measured
  empirically in prior work.
\end{itemize}

The main results of this paper have been verified in the mechanical
proof assistant, Lean. The full mechanization is available in a public
git repository at
\url{https://gitlab.mpi-sws.org/FCS/mixnets-sync-public}.

\paragraph{Structure of the paper}
We present the formal setup, background material on
$(\dpep, \dpde)$-anonymity properties, our threat model, and an
overview of our results in \cref{sec:setup}. \cref{sec:evaluation}
uses the results of earlier sections to explore the trade-off between
anonymity, network latency and per-router bandwidth quantitatively
(with actual numbers). Our actual analysis of the variance leading to
the proofs of our main results is covered in \cref{sec:analysis} and
its subsections.

\section{Setup and Key Results}
\label{sec:setup}

\paragraph{Mixnet setup}
We consider a standard parallel mixnet (henceforth, mixnet) with a set
of servers, $\routers$, of size $\nwidth$. $\routers =
\{\router_1,\ldots,\router_\nwidth\}$. A set of possible senders,
$\senders$, and a set of possible recipients, $\receivers$, form the
users of the system. $\senders$ and $\receivers$ may overlap.

To send a message to a recipient, say $\receiver \in \receivers$, the
message sender, say $\sender \in \senders$, picks a \emph{routing
  path} of $\ndepth > 0$ servers
$\irouter{\rind_1}, \ldots, \irouter{\rind_{\ndepth}}$ by sampling
each $\irouter{\rind_i}$ uniformly and independently at random from
$\routers$ with repetition ($\ndepth$ is a configuration
parameter). The sender \emph{onion encrypts} the message in
$\ndepth+1$ layers using the public keys of the principals
$\receiver, \irouter{\rind_{\ndepth}}, \ldots,
\irouter{\rind_1}$ in that order. (The innermost and
outermost layers of the onion encryption are for $\receiver$ and
$\irouter{\rind_1}$, respectively.) The payload at each layer
also includes the identity of the next principal in the chain. The
sender sends the onion encrypted message to $\irouter{\rind_{1}}$,
the first server on the routing path.

The mixnet's servers operate in loosely synchronized rounds that last
a fixed duration of time. In each round, each server collects messages
that senders or other servers send to it. For each such message, the
server decrypts the outermost layer of encryption (which must be with
the server's own public key), discovering the identity of the
next-in-path node (server or message recipient) to which the payload
must be forwarded. Before forwarding any payload, the server mixes the
round's messages by \emph{shuffling them uniformly at random} by
applying a randomly sampled permutation. At the end of the round, the
server forwards the messages in the shuffled order. A message reaches
its intended recipient after $\ndepth$ rounds of forwarding.

\paragraph{Mixing and anonymity}
The mixnet's operation mixes \emph{all messages a server receives in
  the same round} perfectly uniformly: An adversary with full
visibility of the network, but not of the internal operations of a
server, cannot associate any message entering the server with any
outgoing message from the same round with a likelihood better than a
completely random guess.%
\footnote{For simplicity of exposition, we are describing here the
  specific case without server compromise. Our actual analysis
  accounts for server compromise in accordance with our threat model
  (\cref{sec:threat-model}).}
Thus, between entering the mixnet and exiting it, a message mixes with
$\ndepth$ randomly chosen subsets of messages (the subsets may
overlap). Each mixing increases the confusion for a network adversary
trying to track the message, which leads to anonymity.

However, a parallel mixnet's mixing and anonymity are strictly weaker
than those of an \emph{ideal mixnet}, which consists of exactly one
uncompromised server that mixes \emph{all messages sent concurrently
  during a round} in a single shuffle. In contrast, in a parallel
mixnet, each message mixes with only a randomly chosen $1/\nwidth$
fraction of concurrently sent messages on average in each round. For
large enough $\ndepth$, we expect any message to mix with a
significant fraction of other concurrently sent messages over
$\ndepth$ rounds, but this is a \emph{probabilistic} property, not a
certainty.

We develop a new stochastic analysis to prove that a strong
quantitative anonymity property, $(\dpep, \dpde)$-message
indistinguishability, holds for parallel mixnets and that the property
improves ($\dpep, \dpde$ reduce) with increasing $\nmsgs$ and $\ndepth$,
and decreasing~$\nwidth$, but the relationships are highly non-linear.

\paragraph{Performance}
We consider two performance metrics. 1)~Message latency, the amount of
time needed for a message to be delivered, which we proxy with
$\ndepth$. 2)~Average per-server bandwidth, the average volume of data
every server transmits in every round, which we proxy with
$\nmsgs/\nwidth$.

\subsection{Threat Model}
\label{sec:threat-model}

We consider a mixnet adversary whose objective is to track one or more
messages as they traverse the mixnet. This \emph{adversary is strong
  and honest-but-curious}: It has a full view of all network
communication and it can compromise 
each server independently with probability $\pcompromise$,
partially to obtain their internal shuffling permutations and private
keys, but it does not modify the protocol in any way (else it may be
detected). Our analysis assumes that the mixnet is separated
into $\ndepth$ layers of $\nwidth$ distinct servers each, with each
round using the servers of its own layer, so that whether the server
a message visits in one round is compromised is independent of the
servers it visited in earlier rounds. If instead the same $\nwidth$
servers are reused in every round, the history of a run reveals which
servers are compromised, and our exact results become optimistic: by
Jensen's inequality, the expected squared error of the posterior odds
then exceeds the value we derive in \cref{sec:analysis}. As usual, we assume that the adversary cannot break
cryptography. Attacks using side-channels are beyond the scope of the
standard mixnet model and our analysis.

The adversary may also watermark a subset of messages and track them
from start to finish. This also applies to messages sent by senders
who collude with the adversary. We handle such messages in our
analysis by (conservatively) removing them from the total message
count, $\nmsgs$. This is justified because watermarked messages
provide no privacy to other messages. Thus, sender compromise does not
change our analysis in a significant way; it only changes the value of
one parameter ($\nmsgs$) in our instantiation of the analysis.

Our analysis focuses primarily on the mixing properties of the mixnet
itself. The $(\dpep, \dpde)$-message indistinguishability property
bounds privacy loss from the information the adversary learns by
observing the mixnet routing only. We do not consider any prior
knowledge of the distribution of sender-recipient pairs that the
adversary may have. Our analysis can be extended to cover such
knowledge using (standard) Bayesian techniques.

\subsection{An Abstract Model of Parallel Mixnets}
\label{sec:mixnet-model}

In stating and establishing mixing properties, we find it convenient
to work with an abstract model of parallel mixnets that has an
additional \emph{conceptual} (hypothetical) entity and additional
\emph{conceptual} state.

Define a \emph{cohort} of messages as the set of messages entering the
mixnet in the same round and define a \emph{run} of the mixnet as the
process of the mixnet routing a cohort, all the way from the cohort's
$\nmsgs$ messages being received by the mixnet's $\nwidth$ servers,
getting mixed piecemeal through $\ndepth-1$ forwarding rounds of the
mixnet, and all $\nmsgs$ messages exiting the mixnet towards their
intended recipients.

To simplify our analysis, we assume that there is no mixing of
information across different cohorts. This is a conservative
assumption; in reality, the simultaneous routing of $\ndepth$ cohorts
on the servers confuses the adversary more than what we model.

\paragraph{Abstract model}
Our abstract model is like the real parallel mixnet, except that it
introduces an honest conceptual entity called the \emph{routing
  authority}, $\rauth$, which acts between local shuffle rounds. This
routing authority performs two functions:\\
(1) $\rauth$ takes on the task of deciding the next hop of each
message that was sent by every honest sender --- in other words, the
path of each uncompromised message is determined step-by-step by
$\rauth$ rather than upfront by the message's sender. This does not
change the threat model since only
honest senders' coin tosses are replaced by $\rauth$'s tosses in this way, and \\
(2) $\rauth$ collects all $\nmsgs$ messages of the cohort after each
round of local shuffling, and it shuffles the $\nmsgs$ messages
uniformly at random using a permutation that is \emph{visible to the
  adversary} resulting in an array $\idealarrs{i}$, before sending the
messages to their next hops. Again, this shuffling by $\rauth$ does
not change the threat model because the adversary sees the permutation
that $\rauth$ applies, so it can continue to track each message from a
server to $\rauth$ and from $\rauth$ to the next server with perfect
precision.

Concretely, routing works in the abstract model as follows. Senders of
a cohort send their $\nmsgs$ messages to $\rauth$, which collects the
messages, shuffles them uniformly at random, resulting in an array
$\idealarrs{0}$. $\rauth$ assigns entry point servers to honest
senders' messages (sampling a server uniformly at random for each such
message) and forwards the messages to the respective servers. The
entire operation of $\rauth$, including its permutation, is in plain
sight of the adversary, so it does not affect the adversary's ability
to track messages.

Having received its subset of messages from $\rauth$, each server
locally shuffles the subset using a secret uniformly sampled
permutation (as in the real mixnet), and then forwards the messages
back to $\rauth$ in plain sight of the adversary. $\rauth$ collects
and shuffles all $\nmsgs$ messages, obtaining a new array, called
$\idealarrs{1}$, decides the next hops of honest messages uniformly at
random and forwards all messages to their next hops --- all in plain
sight of the adversary.

This process repeats $\ndepth$ times yielding arrays
$\idealarrs{0}, \idealarrs{1}, \ldots$ until a final array
$\idealarrs{\ndepth}$ is obtained from which messages are forwarded to
their intended recipients.

It should be obvious that this abstract model is functionally
equivalent to the real parallel mixnet in terms of the adversary's
ability to track messages. However, this equivalent system adds global
message arrays
$\idealarrs{0}, \idealarrs{1}, \ldots, \idealarrs{\ndepth}$, whose
elements have been randomly shuffled. We use these arrays to state and
prove mixing properties. Note that the distribution of any attribute
of a message in $\idealarrs{i}$ has to be independent of the message's
\emph{position} in the array, a property we exploit later.

\subsection{Posterior Odds}
\label{sec:forward-probabilities}

Following prior work~\cite{kate25tightrope}, we cast our analysis in terms of
posterior odds, which we attach to all internal messages that arise
during a mixnet run. The posterior odds of a message are the odds that
the message is the partially onion-decrypted variant of a specific
sent message of interest to the adversary, \emph{conditional} on the
adversary's observation of the entire run. We first explain the
observations an adversary makes and then explain posterior odds.

\paragraph{Adversary observation} In our model, the
adversary observes the following information about a run:\\
- The \emph{routing graph}: In each round, the messages sent from each
server to the routing authority, the routing authority's permutation
leading to $\idealarrs{\lind}$, and the messages sent from the routing
authority to their next hops. Due to the onion encryption on messages,
this information is equivalent to knowing for every message leaving a server in a round, which server it
enters in the next round. (When no server is compromised, this is
further equivalent to knowing the number of messages sent between
every pair of servers in every round.)\\
- For compromised servers, the adversary also observes the exact
mapping from the server's incoming messages to its outgoing messages
in each round (the shuffling permutations).

Importantly, the adversary does not learn the shuffling permutations
within honest servers and, therefore, it also does not learn the exact
routing path of each message. It is precisely this hidden information
that provides anonymity.

We use the letter $\obss$ to denote the adversary's observation in a
specific run of interest, and the letter $\obs$ to denote the random
variable which records the adversary's observation. Conditioning on
$\obs = \obss$ means ``under the condition that the adversary observes
$\obss$''.

\begin{figure*}
	\centering
	\begin{tikzpicture}
		[
		roundnode/.style={circle, draw=blue!60, fill=blue!5, very thick, minimum size=7mm},
		squarednode/.style={rectangle, draw=red!60, fill=red!5, very thick, minimum size=5mm},
		lab/.style={fill=white, inner sep=1pt, font=\small, text=black},
		]
		
		\def\vsep{2mm}
		\def\asep{8mm}
		
		\node[squarednode] (u1) at (2,1) {$\msender{1}$};
		\node[squarednode, anchor=south, yshift=\vsep] (u2) at (u1.north) {$\msender{2}$};
		\node[squarednode, anchor=south, yshift=\vsep] (u3) at (u2.north) {$\msender{3}$};
		\node[squarednode, anchor=south, yshift=\vsep] (u4) at (u3.north) {$\msender{4}$};
		\node[squarednode, anchor=south, yshift=\vsep] (u5) at (u4.north) {$\msender{5}$};
		\node[squarednode, anchor=south, yshift=\vsep] (u6) at (u5.north) {$\msender{6}$};
		\node[anchor=south, yshift=\vsep] (L0) at (u6.north) {Senders};
                
		\node[squarednode] (m1) [right=\asep of u1] {$0$};
		\node[squarednode, anchor=south, yshift=\vsep] (m2) at (m1.north) {$1$};
		\node[squarednode, anchor=south, yshift=\vsep] (m3) at (m2.north) {$0$};
		\node[squarednode, anchor=south, yshift=\vsep] (m4) at (m3.north) {$0$};
		\node[squarednode, anchor=south, yshift=\vsep] (m5) at (m4.north) {$0$};
		\node[squarednode, anchor=south, yshift=\vsep] (m6) at (m5.north) {$0$};
		\node[anchor=south, yshift=\vsep] (A0) at (m6.north) {$\idealarrs{0}$};
                
		\node[roundnode] (r11) [right=\asep of m1] {$\irouter{1}$};
		\node[roundnode] (r21) [above=of r11] {$\irouter{2}$};
		\node[roundnode] (r31) [above=of r21] {$\irouter{3}$};
		
		\node[squarednode] (u21) [right=\asep of r11] {$0$};
		\node[squarednode, anchor=south, yshift=\vsep] (u22) at (u21.north) {$\frac{1}{3}$};
		\node[squarednode, anchor=south, yshift=\vsep] (u23) at (u22.north) {$\frac{1}{3}$};
		\node[squarednode, anchor=south, yshift=\vsep] (u24) at (u23.north) {$\frac{1}{3}$};
		\node[squarednode, anchor=south, yshift=\vsep] (u25) at (u24.north) {$0$};
		\node[squarednode, anchor=south, yshift=\vsep] (u26) at (u25.north) {$0$};
		\node[anchor=south, yshift=\vsep] (A1) at (u26.north) {$\idealarrs{1}$};
		
		\node[roundnode] (r12) [right=\asep of u21] {$\irouter{1}$};
		\node[roundnode] (r22) [above=of r12] {$\irouter{2}$};
		\node[roundnode] (r32) [above=of r22] {$\irouter{3}$};
		
		\node[squarednode] (u31) [right=\asep of r12] {$\frac{3}{18}$};
		\node[squarednode, anchor=south, yshift=\vsep] (u32) at (u31.north) {$\frac{3}{18}$};
		\node[squarednode, anchor=south, yshift=\vsep] (u33) at (u32.north) {$\frac{4}{18}$};
		\node[squarednode, anchor=south, yshift=\vsep] (u34) at (u33.north) {$\frac{4}{18}$};
		\node[squarednode, anchor=south, yshift=\vsep] (u35) at (u34.north) {$\frac{4}{18}$};
		\node[squarednode, anchor=south, yshift=\vsep] (u36) at (u35.north) {$0$};
		\node[anchor=south, yshift=\vsep] (A2) at (u36.north) {$\idealarrs{2}$};
		
		\node[roundnode] (r13) [right=\asep of u31] {$\irouter{1}$};
		\node[roundnode] (r23) [above=of r13] {$\irouter{2}$};
		\node[roundnode] (r33) [above=of r23] {$\irouter{3}$};
		
		\node[squarednode] (u41) [right=\asep of r13] {$\frac{18}{108}$};
		\node[squarednode, anchor=south, yshift=\vsep] (u42) at (u41.north) {$\frac{21}{108}$};
		\node[squarednode, anchor=south, yshift=\vsep] (u43) at (u42.north) {$\frac{21}{108}$};
		\node[squarednode, anchor=south, yshift=\vsep] (u44) at (u43.north) {$\frac{16}{108}$};
		\node[squarednode, anchor=south, yshift=\vsep] (u45) at (u44.north) {$\frac{16}{108}$};
		\node[squarednode, anchor=south, yshift=\vsep] (u46) at (u45.north) {$\frac{16}{108}$};
                \node[anchor=south, yshift=\vsep] (A3) at (u46.north) {$\idealarrs{3}$};

		\node[squarednode] (r1) [right= of u41] {$\receiver{1}$};
		\node[squarednode, anchor=south, yshift=\vsep] (r2) at (r1.north) {$\receiver{2}$};
		\node[squarednode, anchor=south, yshift=\vsep] (r3) at (r2.north) {$\receiver{3}$};
		\node[squarednode, anchor=south, yshift=\vsep] (r4) at (r3.north) {$\receiver{4}$};
		\node[squarednode, anchor=south, yshift=\vsep] (r5) at (r4.north) {$\receiver{5}$};
		\node[squarednode, anchor=south, yshift=\vsep] (r6) at (r5.north) {$\receiver{6}$};
		\node[anchor=south, yshift=\vsep] (L1) at (r6.north) {Recipients};
		
		\draw[thick,->] (u1.east) -- node[pos=.55,below,lab]{$0$} (m1.west);
		\draw[thick,->] (u2.east) -- node[pos=.55,below,lab]{$1$} (m2.west);
		\draw[thick,->] (u3.east) -- node[pos=.55,below,lab]{$0$} (m3.west);
		\draw[thick,->] (u4.east) -- node[pos=.55,below,lab]{$0$} (m4.west);
		\draw[thick,->] (u5.east) -- node[pos=.55,below,lab]{$0$} (m5.west);
		\draw[thick,->] (u6.east) -- node[pos=.55,below,lab]{$0$} (m6.west);
		
		\draw[thick,->] (m1.east) -- node[pos=.55,below,lab]{$0$} (r11.west);
		\draw[thick,->] (m2.east) -- node[pos=.55,below,lab]{$1$} (r21.west);
		
		\draw[thick,->] (m3.east) to[bend left=18]
		node[pos=.55,below,lab]{$0$} (r21.west);
		
		\draw[thick,->] (m4.east) to[bend right=18]
		node[pos=.55,above,lab]{$0$} (r21.west);
		
		\draw[thick,->] (m5.east) -- node[pos=.55,below,lab]{$0$} (r31.west);
		
		\draw[thick,->] (m6.east) to[bend right=15]
		node[pos=.55,above,lab]{$0$} (r31.west);
		
		\draw[thick,->] (r11.east) -- node[pos=.55,below,lab]{$0$} (u21.west);
		
		\draw[thick,->] (r21.east) to[bend left=18]
		node[pos=.65,below,lab]{$\frac{1}{3}$} (u22.west);
		
		\draw[thick,->] (r21.east) --
		node[pos=.55,lab]{$\frac{1}{3}$} (u23.west);
		
		\draw[thick,->] (r21.east) to[bend right=18]
		node[pos=.55,above,lab]{$\frac{1}{3}$} (u24.west);
		
		\draw[thick,->] (r31.east) --
		node[pos=.55,below,lab]{$0$} (u25.west);
		
		\draw[thick,->] (r31.east) to[bend right=15]
		node[pos=.55,above,lab]{$0$} (u26.west);
		
		\draw[thick,->] (u21.east) to[bend left=15]
		node[pos=.55,below,lab]{$0$} (r12.west);
		
		\draw[thick,->] (u22.east) to[bend right=15]
		node[pos=.55,above,lab]{$\frac{1}{3}$} (r12.west);
		
		\draw[thick,->] (u23.east) to[bend left=15]
		node[pos=.55,below,lab]{$\frac{1}{3}$} (r22.west);
		
		\draw[thick,->] (u24.east) to[bend right=15]
		node[pos=.55,above,lab]{$\frac{1}{3}$} (r22.west);
		
		\draw[thick,->] (u25.east) --
		node[pos=.55,below,lab]{$0$} (r32.west);
		
		\draw[thick,->] (u26.east) to[bend right=12]
		node[pos=.55,above,lab]{$0$} (r32.west);
		
		\draw[thick,->] (r12.east) -- node[pos=.55,below,lab]{$\frac{3}{18}$} (u31.west);
		\draw[thick,->] (r12.east) to[bend left=18]
		node[pos=.65,below,lab]{$\frac{3}{18}$} (u32.west);

		\draw[thick,->] (r22.east) --
		node[pos=.55,below,lab]{$\frac{4}{18}$} (u33.west);
		\draw[thick,->] (r22.east) to[bend right=18]
		node[pos=.55,lab]{$\frac{4}{18}$} (u34.west);
		\draw[thick,->] (r22.east) --
		node[pos=.55,above,lab]{$\frac{4}{18}$} (u35.west);

		\draw[thick,->] (r32.east) to[bend right=15]
		node[pos=.55,above,lab]{$0$} (u36.west);
		
		\draw[thick,->] (u31.east) to[bend left=15]
		node[pos=.55,below,lab]{$\frac{3}{18}$} (r13.west);
		
		\draw[thick,->] (u32.east) to[bend right=15]
		node[pos=.55,below,lab]{$\frac{3}{18}$} (r23.west);
		\draw[thick,->] (u33.east) to[bend left=15]
		node[pos=.55,lab]{$\frac{4}{18}$} (r23.west);
		
		\draw[thick,->] (u34.east) to[bend right=15]
		node[pos=.55,below,lab]{$\frac{4}{18}$} (r33.west);
		\draw[thick,->] (u35.east) --
		node[pos=.35,lab]{$\frac{3}{18}$} (r33.west);	
		\draw[thick,->] (u36.east) to[bend right=12]
		node[pos=.55,above,lab]{$0$} (r33.west);
		
		\draw[thick,->] (r13.east) -- node[pos=.55,below,lab]{$\frac{18}{108}$} (u41.west);
		
		\draw[thick,->] (r23.east) to[bend left=18]
		node[pos=.4,below,lab]{$\frac{21}{108}$} (u42.west);
		\draw[thick,->] (r23.east) --
		node[pos=.55,above,lab]{$\frac{21}{108}$} (u43.west);
		
		\draw[thick,->] (r33.east) to[bend right=18]
		node[pos=.55,below,lab]{$\frac{16}{108}$} (u44.west);
		\draw[thick,->] (r33.east) --
		node[pos=.45,lab]{$\frac{16}{108}$} (u45.west);
		\draw[thick,->] (r33.east) to[bend right=15]
		node[pos=.45,above,lab]{$\frac{16}{108}$} (u46.west);
		
		\draw[thick,->] (u41.east) -- node[pos=.55,below,lab]{$\frac{18}{108}$} (r1.west);
		\draw[thick,->] (u42.east) -- node[pos=.55,below,lab]{$\frac{21}{108}$} (r2.west);
		\draw[thick,->] (u43.east) -- node[pos=.55,below,lab]{$\frac{21}{108}$} (r3.west);
		\draw[thick,->] (u44.east) -- node[pos=.55,below,lab]{$\frac{16}{108}$} (r4.west);
		\draw[thick,->] (u45.east) -- node[pos=.55,below,lab]{$\frac{16}{108}$} (r5.west);
		\draw[thick,->] (u46.east) -- node[pos=.55,below,lab]{$\frac{16}{108}$} (r6.west);
		
	\end{tikzpicture}
	\caption{An illustrative calculation of posterior odds. Here,
          all three servers, $\irouter{1}$, $\irouter{2}$,
          $\irouter{3}$ are assumed to be honest, and the averaging
          rule is applied for each forwarding. Observe how the highly
          skewed starting odds $[0,1,0,0,0,0]$ diffuse into nearly
          uniform odds
          $\left[\frac{18}{108}, \frac{21}{108}, \frac{21}{108},
            \frac{16}{108}, \frac{16}{108}, \frac{16}{108}\right]$ in
          just three rounds of mixing.}
\label{fig:forward-probs-abstract}
\end{figure*}
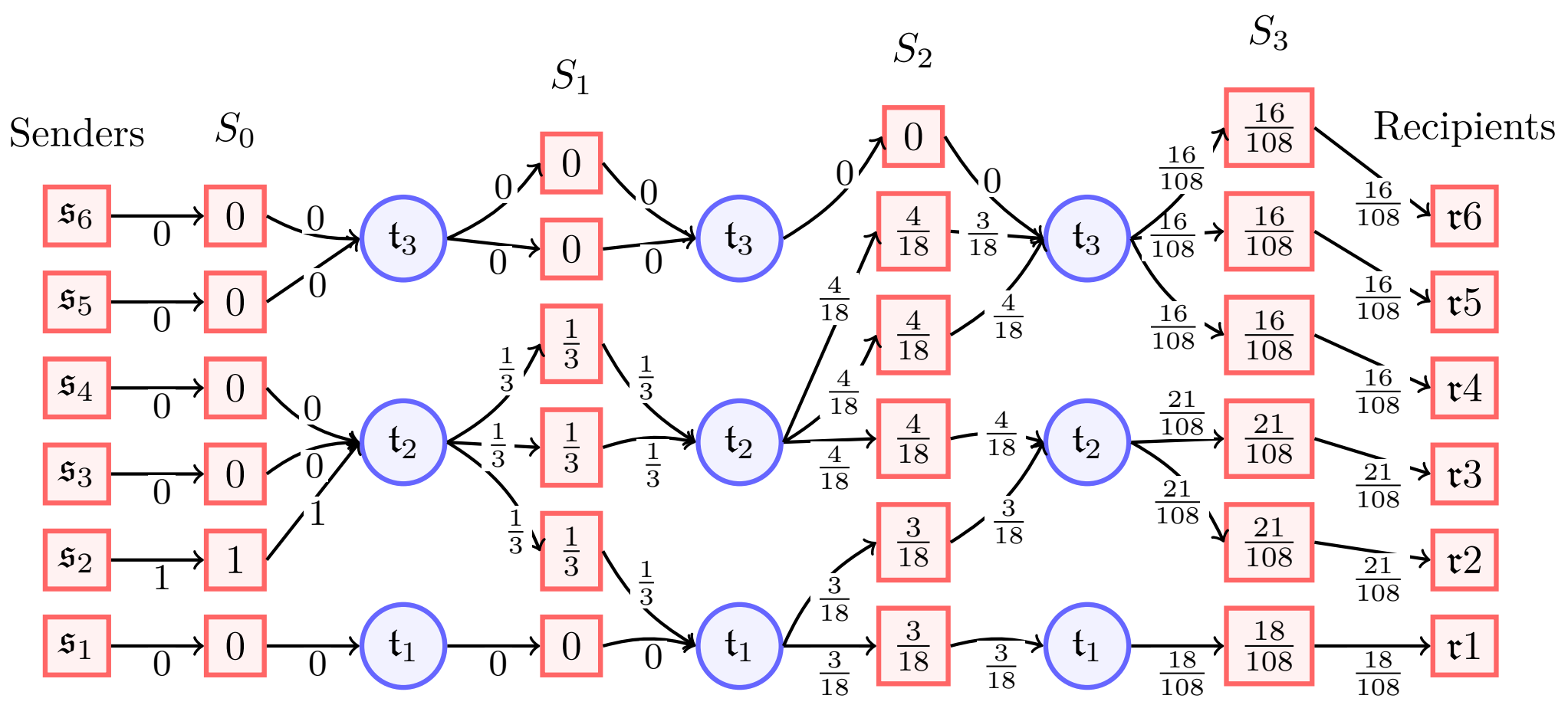

\paragraph{Posterior odds}
Given its observation about a run, the adversary can calculate the
posterior odds layer-by-layer, i.e., iteratively. It is convenient to
associate these odds with elements of the arrays
$\idealarrs{0}, \idealarrs{1}, \ldots, \idealarrs{\ndepth}$:\\
- $\idealarrs{0}$: This is the entry point of the run's cohort. Here,
the posterior odds are $1$ for the message of interest to the
adversary, and $0$ everywhere else. \\
- $\idealarrs{\lind+1}$: To calculate the posterior odds for
this array, repeat the following for every server $\irouter{\rind}$:
\begin{itemize}
\item If $\irouter{\rind}$ is honest, take from $\idealarrs{\lind}$
  the posterior odds of all messages that were mixed by
  $\irouter{\rind}$ in round $\lind+1$. \emph{Average} these forward
  odds. Assign the average odds to all messages in
  $\idealarrs{\lind+1}$ that were sent out by $\irouter{\rind}$.
\item If $\irouter{\rind}$ is dishonest, its shuffling permutation is
  known to the adversary, so its incoming and outgoing messages can be
  mapped one-to-one. Copy the posterior odds of each incoming
  message from $\idealarrs{\lind}$ to the corresponding message in
  $\idealarrs{\lind+1}$.
\end{itemize}
The rationale for assigning the same average odds to all the outgoing
messages of an honest server is that the adversary cannot determine
the mapping from incoming to outgoing messages with any certainty
better than a random guess.

\cref{fig:forward-probs-abstract} shows this iterative process
graphically for a network with three servers that are all honest. The
first layer consists of senders $\sender_1,\ldots,\sender_6$. The
adversary is interested in tracking the message of $\sender_2$, so we
mark that message with posterior odds $1$ in the first array
$\idealarrs{0}$, and all other messages with posterior odds $0$. In
subsequent layers, we assign odds following the averaging rule.

Observe how the completely skewed posterior odds $[0,1,0,0,0,0]$ of
$\idealarrs{0}$ smoothen to the nearly uniform odds
$\left[\frac{18}{108}, \frac{21}{108}, \frac{21}{108}, \frac{16}{108},
  \frac{16}{108}, \frac{16}{108}\right]$ of $\idealarrs{3}$ in just
three rounds of mixing. It is exactly this rapid mixing property of
parallel mixnets that we analyze quantitatively.

\paragraph{Posterior odds as a random variable, $\postpr{\lind}{\mind}$}
Let $\postpr{\lind}{\mind}$ denote the posterior odds associated with
the message $\idealarrs{\lind}[\mind]$ for a given run. Given a fixed
message of interest to the adversary, $\postpr{\lind}{\mind}$ is a
\emph{function} of the adversary's observation $\obss$, which itself
depends on the randomness in chosen paths, servers compromised and
shuffling permutations of compromised servers. Hence, in technical
terms, for any $\lind, \mind$, the posterior odds
$\postpr{\lind}{\mind}$ is a \emph{random variable} over all
randomness in the system, except the randomness of onion encryption by
honest senders and the shuffling permutations of servers.

\paragraph{Posterior odds represent actual probabilities}
The calculated posterior odds $\postpr{\lind}{\mind}$ is the actual
\emph{probability} that the given sent message was the message at
position $\idealarrs{\lind}[\mind]$, conditional on the adversary's
observation (the routing graph of the run). Let
$\msg_1 \approx \msg_2$ mean that $\msg_1$ is obtained by onion
decryption of some layers of $\msg_2$.

\begin{lemma}\label{lem:odds-are-probs}
  For a given adversary observation $\obss$ and a given message of
  interest $\msg$, $\postpr{\lind}{\mind}$ is exactly the conditional
  probability
  $\prd{\partinf}{\pcond{\idealarrs{\lind}[\mind] \approx \msg}{\obs =
      \obss}}$, where $\partinf$ denotes the randomness used in the
  operation of the mixnet. Note that this probability is over all the
  sources of randomness including the randomness that determines the
  shuffling permutations of honest servers and the routing paths of
  messages from honest senders.
\end{lemma}
\begin{proof}
  By straightforward induction on rounds, $\lind$.
\end{proof}

\subsection{Anonymity Property: $(\dpep, \dpde)$-Message Indistinguishability}
\label{sec:anonymity-property}

The strong anonymity property we prove is based on prior
work~\cite{kate13anoa,kate25tightrope}. The property can be defined as
follows in our notation.

\begin{defn}\label{def:message-indistinguishability}
  A parallel mixnet operating with parameters $\nmsgs$, $\nwidth$,
  $\ndepth$, $\pcompromise$ has $(\dpep, \dpde)$-message
  indistinguishability if for any two distinct indices $\mind_1, \mind_2$ and
  any sent message $\msg$ of interest to the adversary,
  the following holds with probability at least $1 - \dpde$
  over $f \sim \partinf$:
  \[\begin{array}{l}
    \prd{\partinf}{\pcond{\idealarrs{\ndepth}[\mind_1]
    \approx \msg}{\obs = \obss_{f}}} \\
    \qquad\qquad \leq e^\dpep \cdot \prd{\partinf}{\pcond{\idealarrs{\ndepth}[\mind_2]
    \approx \msg}{\obs = \obss_{f}}}
  \end{array}\]
Here, $\partinf$ denotes the randomness used in the operation of the
mixnet, $f$ is a sample drawn from $\partinf$, and $\obss_{f}$ denotes
the observation made by the adversary for the run under this
sample. $(\dpep, \dpde)$ may depend on $\nmsgs$, $\nwidth$, $\ndepth$,
$\pcompromise$.
\end{defn}

Intuitively, the two conditional probabilities are the
adversary's beliefs, after observing the run $f$, that the message of
interest exits at position $\mind_1$ and at position $\mind_2$. The
definition says that, except on runs of total probability at most
$\dpde$, these beliefs differ by a factor of at most $e^\dpep$. This is
the \emph{probabilistic} form of differential privacy, which is
stronger than the more common approximate form. The quantifier over runs is essential. If we averaged the
beliefs over all runs before comparing them, both would equal
$1/\nmsgs$ (\cref{thm:prob-mean}), and the definition would hold
vacuously. If we instead required the inequality on every run, no
mixnet with $\pcompromise > 0$ could satisfy it: with probability
$\pcompromise^\ndepth$ the message of interest visits only
compromised servers, and the adversary then knows exactly where it
exits.

\section{Key Results}
\label{sec:key-results}

This section presents our key theoretical results. We defer the
detailed analysis and the proofs of theorems covered here to
\cref{sec:analysis}.

\paragraph{Variance of $\postpr{\lind}{\mind}$}
Our first result, which forms the basis for all other results, is a
closed-form formula for the variance of each
$\postpr{\lind}{\mind}$. This variance is a function of the forwarding
round $\lind$. In fact, it reduces exponentially in $\lind$.
(Note that the mean of each $\postpr{\lind}{\mind}$ is exactly
$1/\nmsgs$ because $\sum_{\mind = 1}^\nmsgs \postpr{\lind}{\mind}$ and
all $(\postpr{\lind}{\mind})_{\mind=1}^\nmsgs$ are identically
distributed due to the random shuffling of $\idealarrs{\lind}$ in our
abstract model.)

\begin{thm}\label{thm:prob-var}
  For any $0\leq\lind\leq\ndepth$ and
  $1 \leq \mind \leq \nmsgs$,
	\[
	\mathrm{Var}_{\partinf}\left[\postpr{\lind}{\mind}\right] =
	\beta^\lind \cdot \frac{1}{\nmsgs}\left(1-\frac{1}{\nmsgs}\right)
	\]
	where $\beta = \contfact{\nmsgs}{\nwidth} \cdot \left(1 - \alpha\right) + \alpha$
	and $\contfact{\nmsgs}{\nwidth}$ is the following constant
	independent of $\lind$:
	\[
	\contfact{\nmsgs}{\nwidth} = \frac{\nwidth-1}{\nmsgs-1}\left[ 1 - \left(1 - \frac{1}{\nwidth} \right)^{\nmsgs-1} \right] < \frac{\nwidth}{\nmsgs}
	\]
\end{thm}

Observe that $\beta$ is the sum of two quantities: The probability of
server compromise, $\pcompromise$, and a term which falls nearly
inversely with $\nwidth$. Consequently, when $\pcompromise$ is nearly
zero (no server compromise), the variance of $\postpr{\lind}{\mind}$
falls rapidly as either $\nmsgs$ or $\lind$ increases. As that
variance approaches $0$, each $\postpr{\lind}{\mind}$ becomes the
point distribution at $1/\nmsgs$, which corresponds to the anonymity
of the ideal mixnet. This is why parallel mixnets provide excellent
anonymity when $\pcompromise$ is low and $\nmsgs$ is high, even with
a small number of mixing rounds.

When $\pcompromise$ is non-negligible, it dominates $\beta$ and
increasing $\nwidth$ does not lower the variance substantially beyond
a point. This is expected: With high $\pcompromise$, the chances that
the message being tracked lands only on compromised servers increase
(in fact, this chance is exactly $\pcompromise^\ndepth$) and this
failure probability dominates the other failure probability where the
tracked message ends up on servers with few other messages.

\paragraph{Tail bound on $\postpr{\lind}{\mind}$}
Using a concentration bound (specifically, Chebyshev's inequality) on
\cref{thm:prob-var}, we can immediately calculate a bound on the
probability that any $\postpr{\lind}{\mind}$ deviates more than $a$
from the mean, $1/\nmsgs$.

\begin{cor}[Tail bounds on $\postpr{\lind}{\mind}$]\label{cor:tail-bound-expec1}  
  For any $\msg$ position $\mind$ and any constant
  $0 < a < \frac{1}{\nmsgs}$,
	\[ \begin{aligned}
	\prd{\partinf}{ \; \left| \, \postpr{\lind}{\mind}-\frac{1}{\nmsgs} \, \right| \; \geq a}
	&\leq
	\beta^{\lind} \frac{1}{\nmsgs a^2} \left(1 - \frac{1}{\nmsgs}\right) \\
	&<
	\beta^{\lind} \frac{1}{\nmsgs a^2}
	\end{aligned} \]
      for the $\beta$ defined in \cref{thm:prob-var}.
\end{cor}

\paragraph{Indistinguishability of exiting messages}
We now state our main theorem (technically, a corollary) --- a
parallel mixnet satisfies $(\dpep,\dpde)$-message indistinguishability
(\cref{def:message-indistinguishability}). The corollary's statement
includes a precise relation between the mixnet's parameters, $\dpep$
and $\dpde$. In practical terms, this result is a statement about the
indistinguishability of exiting messages from each other.

\begin{cor}[$(\dpep,\dpde)$-message indistinguishability]
\label{cor:indis_exit_msgs}
  A parallel mixnet has $(\dpep,\dpde)$-message indistinguishability
  for any $\dpep, \dpde$ satisfying
  \[
    \dpde \geq 2 \beta^{\ell} \cdot \left(\nmsgs - 1\right) \cdot \left[ 1 + \frac{\nmsgs}{2 \left( \nmsgs - 1\right) \sinh^2 \left(\frac{\varepsilon}{2}\right)}\right]
  \]
  where $\beta$ is as defined in \cref{thm:prob-var} and $\sinh(x)$ is
  the standard hyperbolic sine function defined as
  $\sinh(x) \defeq (e^{x}-e^{-x})/2$.
\end{cor}

The relationship between $\dpep$ and $\dpde$ can be inverted to the
following form. Here, $\arsinh$ is the inverse of the hyperbolic sine:
$\arsinh(x) \defeq \ln\left(x + \sqrt{x^2+1}\right)$.
\begin{equation}
  \dpep \geq 2 \arsinh \left( \sqrt{\frac{N}{\frac{\dpde}{\beta^\ndepth} - 2(\nmsgs-1)}} \right)
  \label{eqn:eps-del}
\end{equation}

We show the practical consequences of this corollary by exploring
possible values of $\dpep, \dpde$ for reasonable values of the
operating parameters $\nwidth$, $\nmsgs$, $\ndepth$ and $\pcompromise$
in \cref{sec:evaluation}.

\paragraph{Indistinguishability of entering messages}
Remarkably, our entire analysis can also be run \emph{backwards} from
the exiting messages to the entering messages, to obtain results for
indistinguishability of messages that enter the system, provably
establishing that parallel mixnets also prevent the adversary from
determining with confidence who the sender of an exiting message
is. The privacy parameters $\dpep, \dpde$ we get for this
indistinguishability are exactly the same as for the forwards analysis
(\cref{cor:indis_exit_msgs} above), so we do not write these reverse
results here formally.

\section{Provable Anonymity-Performance Trade-off in Practice}
\label{sec:evaluation}

\begin{figure*}[tp]%
	\centering%
	\begin{subfigure}{0.40\textwidth}
		\centering
		\includegraphics[width=\linewidth]{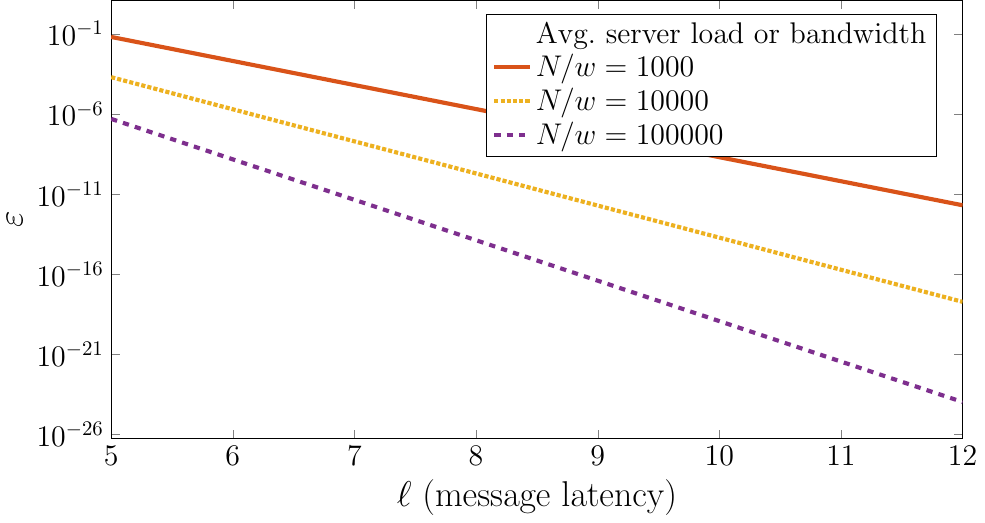}
		\caption{$\alpha = 0$ and $\nmsgs = 10^6$.}
		\label{fig:SH_6H}
	\end{subfigure}
	\begin{subfigure}{0.40\textwidth}
		\centering
		\includegraphics[width=\linewidth]{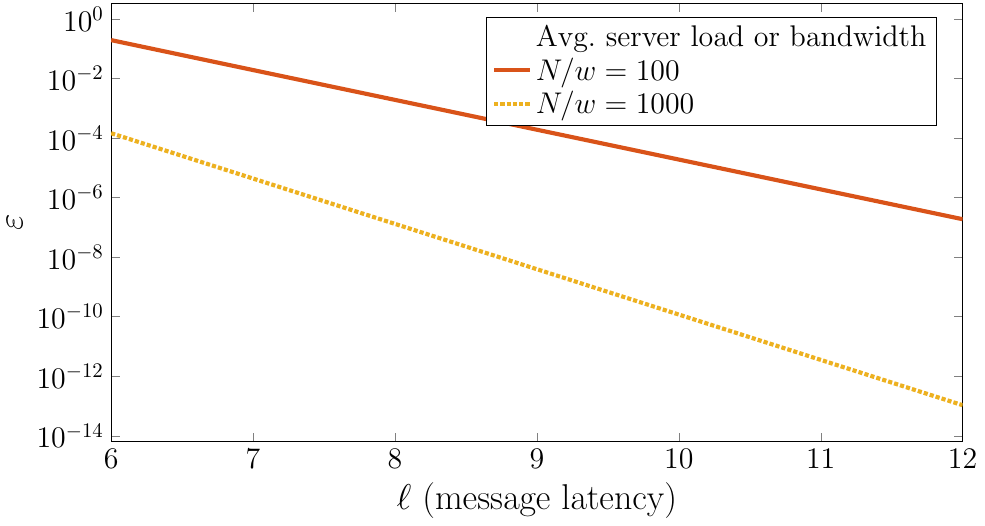}
		\caption{$\alpha = 0$ and $\nmsgs = 10^4$.}
		\label{fig:SH_6L}
	\end{subfigure}
	\centering%
	
	\vspace{1em}%
	\begin{subfigure}{0.40\textwidth}
		\centering
		\includegraphics[width=\linewidth]{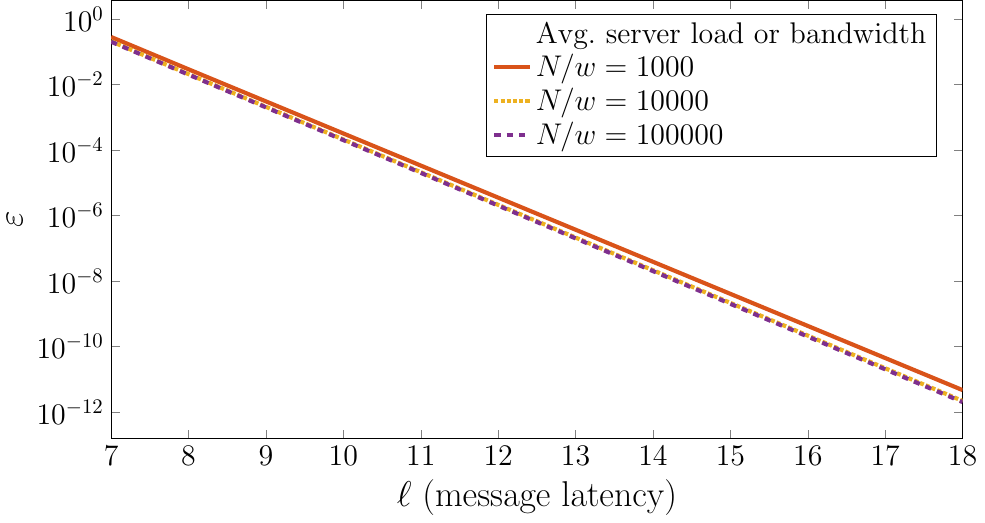}%
		\caption{$\alpha = 0.01$ and $\nmsgs = 10^6$.}
		\label{fig:M_1_6H}%
	\end{subfigure}%
	\begin{subfigure}{0.40\textwidth}
		\centering
		\includegraphics[width=\linewidth]{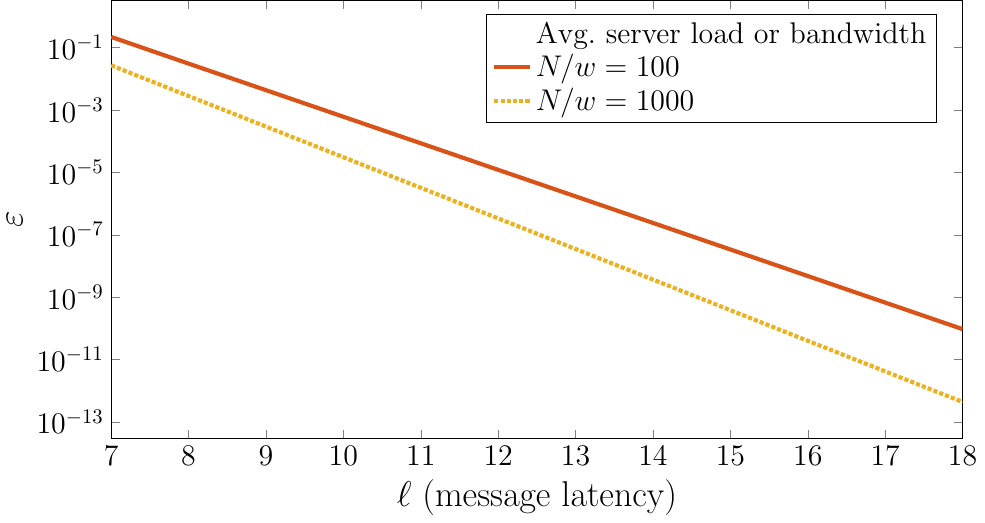}%
		\caption{$\alpha = 0.01$ and $\nmsgs = 10^4$.}
		\label{fig:M_1_6L}%
	\end{subfigure}%
	\centering%
	
	\vspace{1em}%
	\begin{subfigure}{0.40\textwidth}
		\centering
		\includegraphics[width=\linewidth]{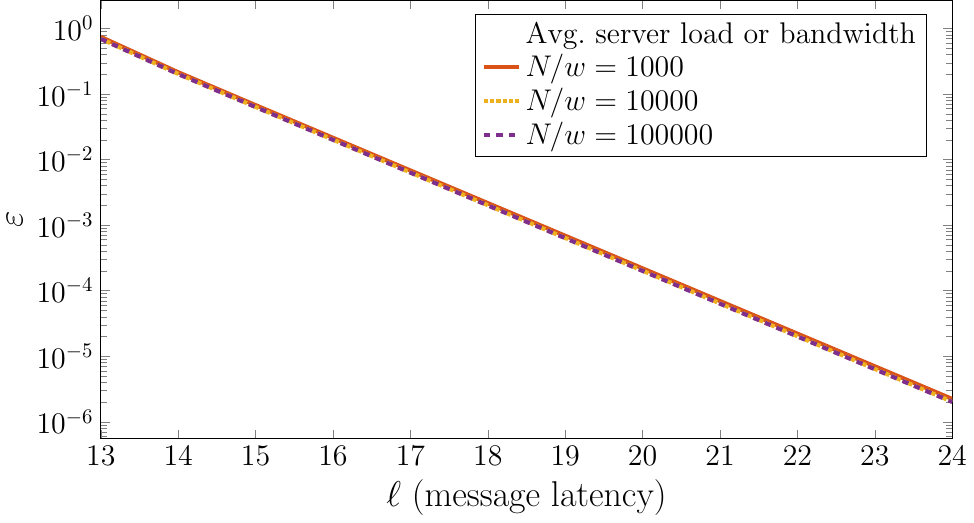}%
		\caption{$\alpha = 0.1$ and $\nmsgs = 10^6$.}
		\label{fig:M_10_6H}%
	\end{subfigure}%
	\begin{subfigure}{0.40\textwidth}
		\centering
		\includegraphics[width=\linewidth]{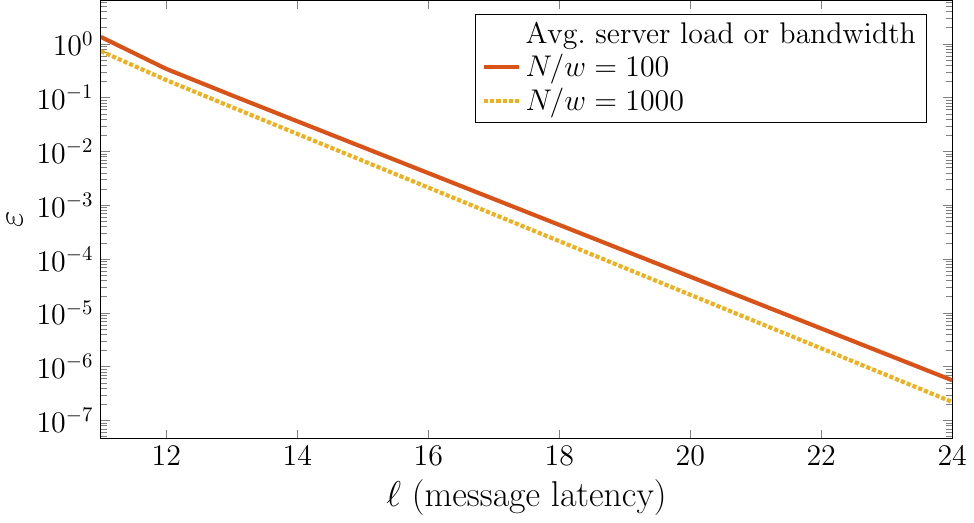}%
		\caption{$\alpha = 0.1$ and $\nmsgs = 10^4$.}
		\label{fig:M_10_6L}%
	\end{subfigure}%
	\centering%
	
	\vspace{1em}%
	\begin{subfigure}{0.40\textwidth}
		\centering
		\includegraphics[width=\linewidth]{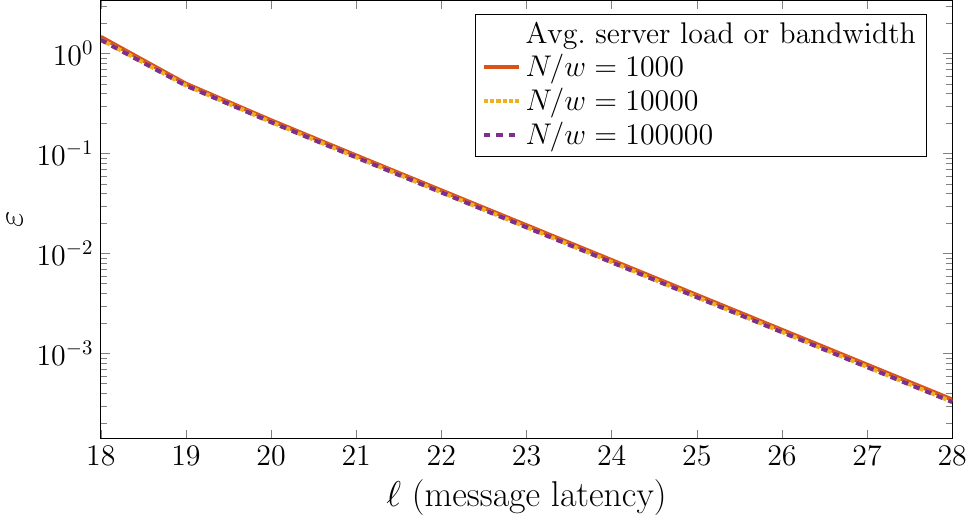}%
		\caption{$\alpha = 0.2$ and $\nmsgs = 10^6$.}
		\label{fig:M_20_6H}%
	\end{subfigure}%
	\begin{subfigure}{0.40\textwidth}
		\centering
		\includegraphics[width=\linewidth]{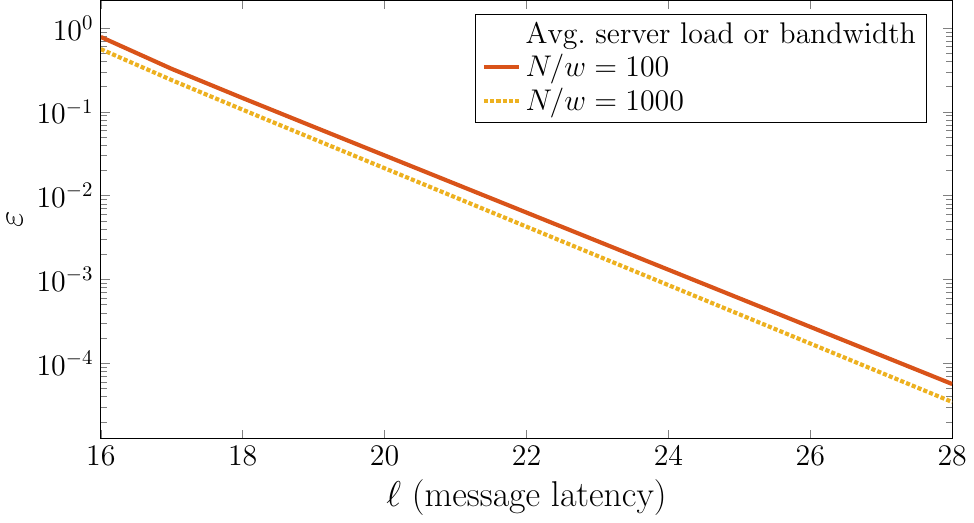}%
		\caption{$\alpha = 0.2$ and $\nmsgs = 10^4$.}
		\label{fig:M_20_6L}%
	\end{subfigure}%
	\centering%
	\caption{Provable minimum values of $\dpep$ (on a logarithmic scale) for varying $\pcompromise$, $\ndepth$ and $\frac{\nmsgs}{\nwidth}$. $\nmsgs$ is held fixed at either $10^6$ for the figures on the left or at $10^4$ for the figures on the right. $\dpde$ is fixed at $10^{-6}$.}%
	\label{fig:epsi_delta_delta6}%
\end{figure*}

\begin{figure*}[tp]%
	\centering%
	\begin{subfigure}{0.40\textwidth}
		\centering
		\includegraphics[width=\linewidth]{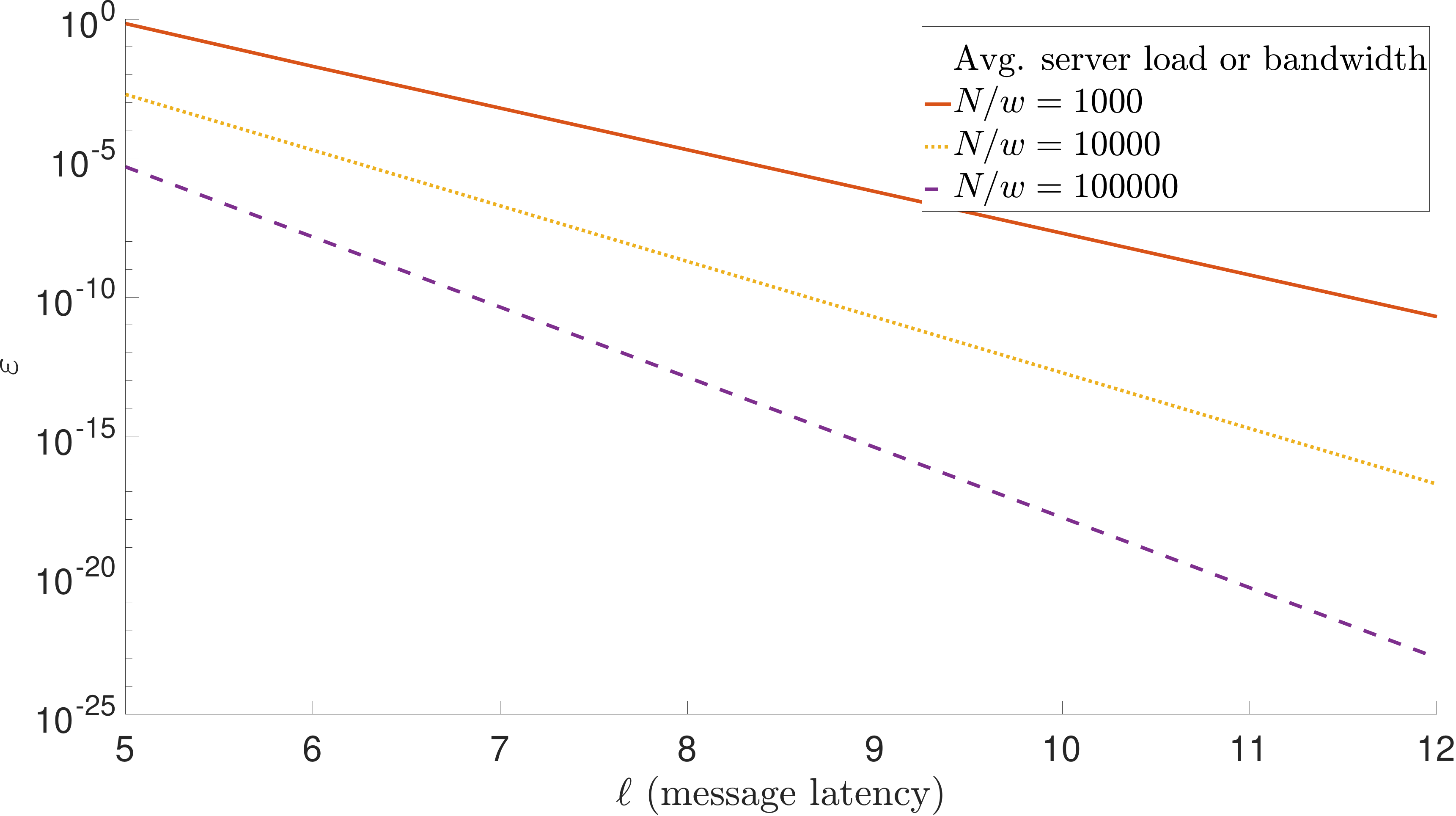}
		\caption{$\alpha = 0$ and $\nmsgs = 10^6$.}
		\label{fig:SH_8H}
	\end{subfigure}
	\begin{subfigure}{0.40\textwidth}
		\centering
		\includegraphics[width=\linewidth]{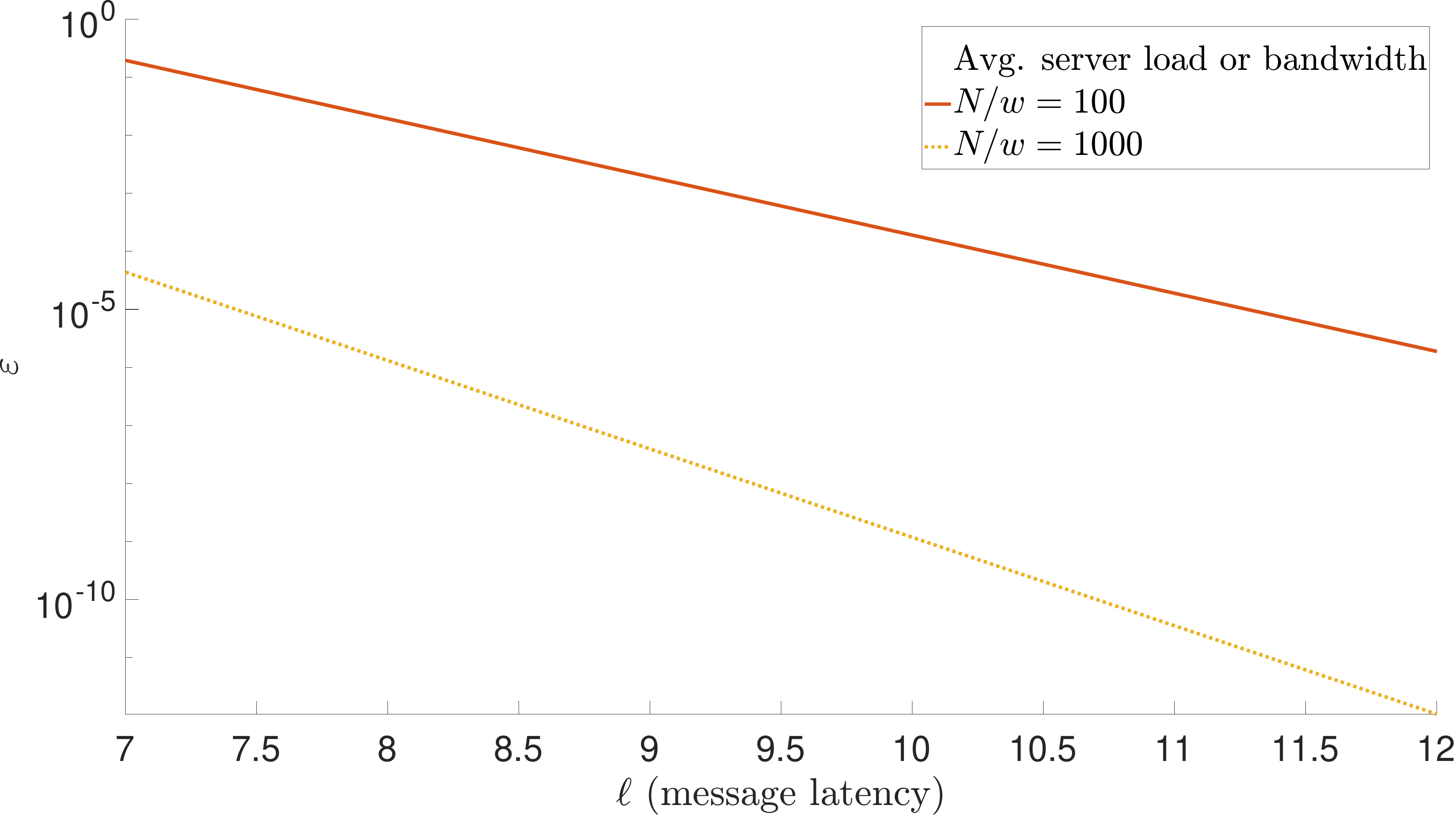}
		\caption{$\alpha = 0$ and $\nmsgs = 10^4$.}
		\label{fig:SH_8L}
	\end{subfigure}
	\centering%
	
	\begin{subfigure}{0.40\textwidth}
		\centering
		\includegraphics[width=\linewidth]{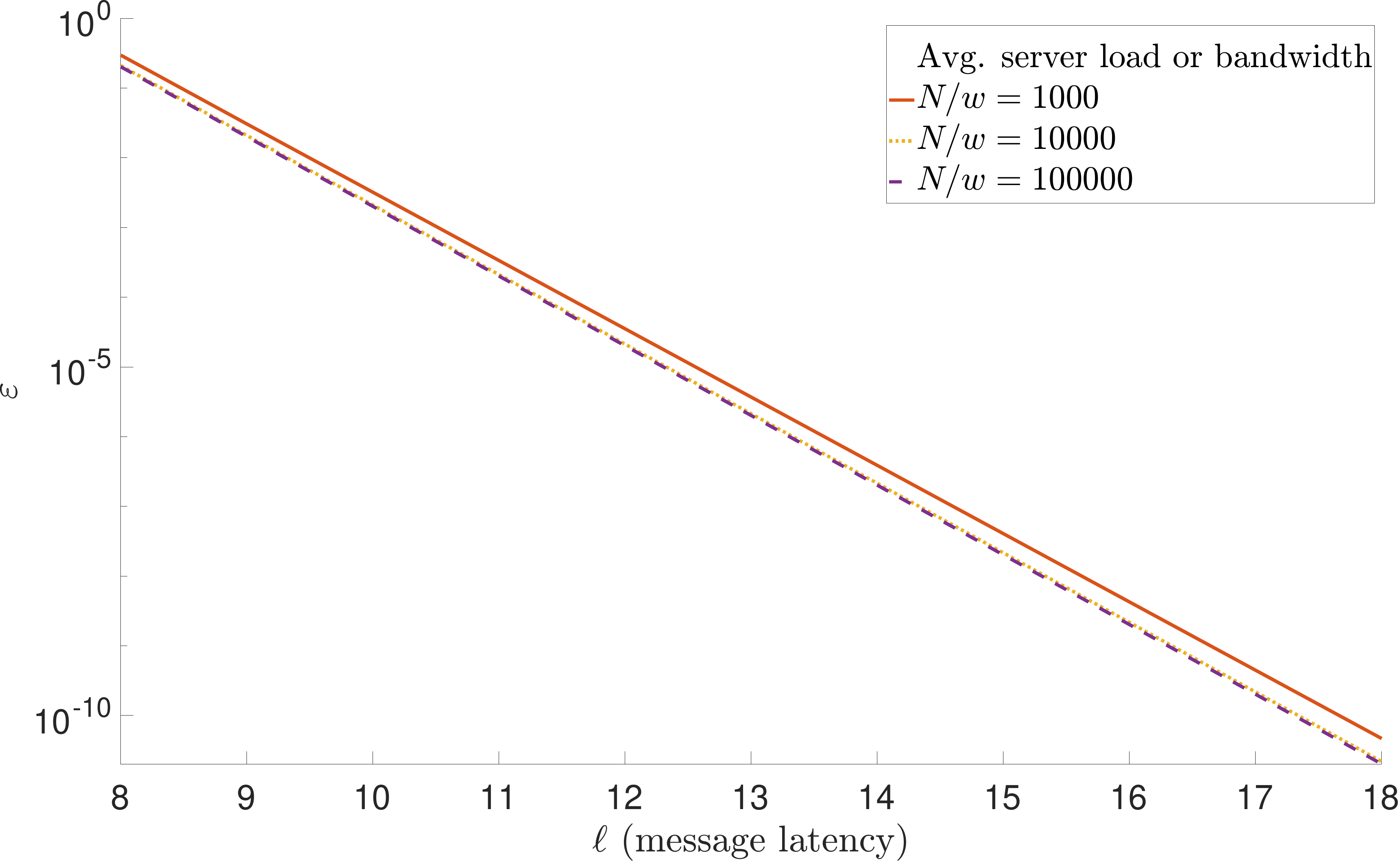}%
		\caption{$\alpha = 0.01$ and $\nmsgs = 10^6$.}
		\label{fig:M_1_8H}%
	\end{subfigure}%
	\begin{subfigure}{0.40\textwidth}
		\centering
		\includegraphics[width=\linewidth]{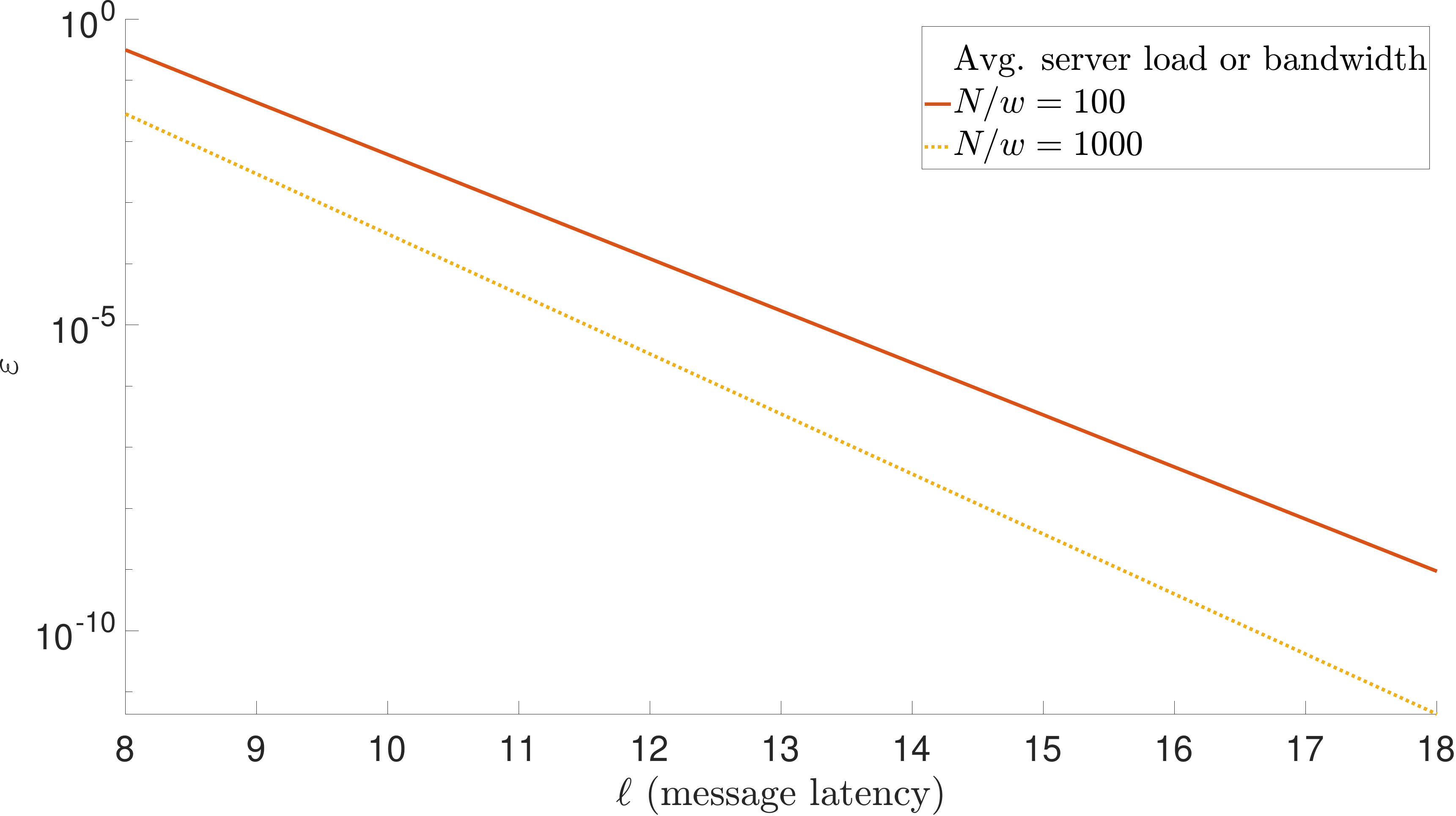}%
		\caption{$\alpha = 0.01$ and $\nmsgs = 10^4$.}
		\label{fig:M_1_8L}%
	\end{subfigure}%
	\centering%
	
	\begin{subfigure}{0.40\textwidth}
		\centering
		\includegraphics[width=\linewidth]{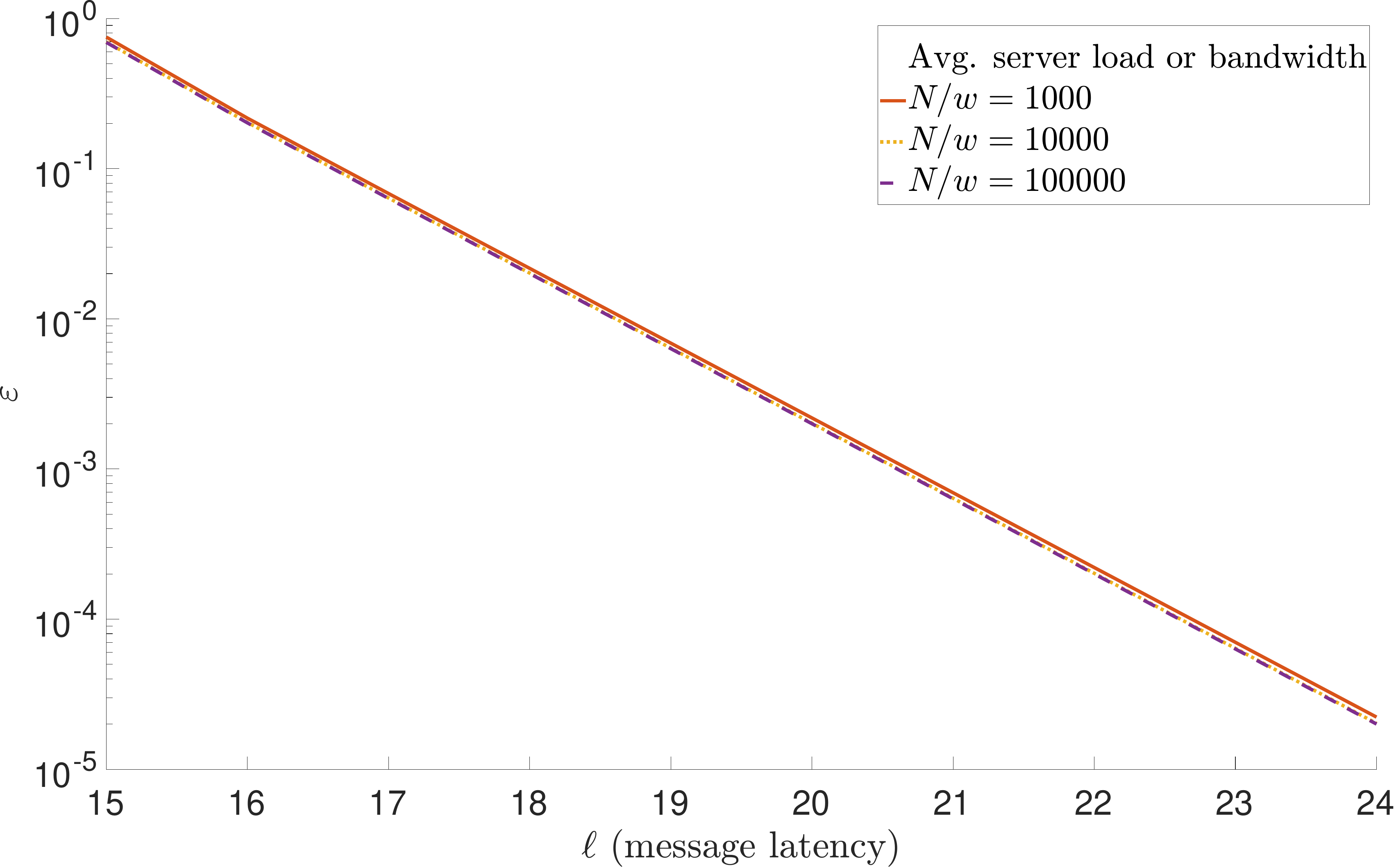}%
		\caption{$\alpha = 0.1$ and $\nmsgs = 10^6$.}
		\label{fig:M_10_8H}%
	\end{subfigure}%
	\begin{subfigure}{0.40\textwidth}
		\centering
		\includegraphics[width=\linewidth]{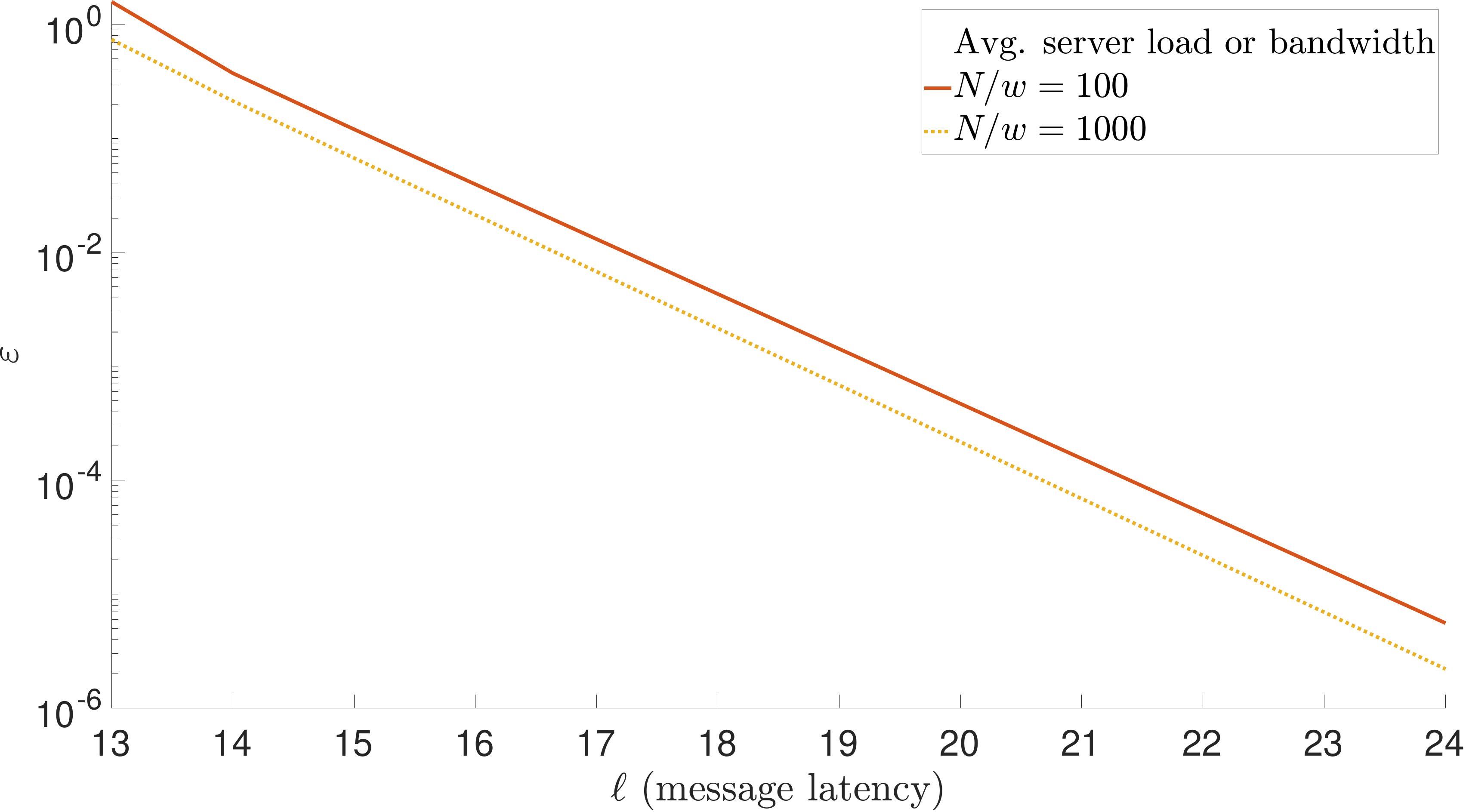}%
		\caption{$\alpha = 0.1$ and $\nmsgs = 10^4$.}
		\label{fig:M_10_8L}%
	\end{subfigure}%
	\centering%
	
	\begin{subfigure}{0.40\textwidth}
		\centering
		\includegraphics[width=\linewidth]{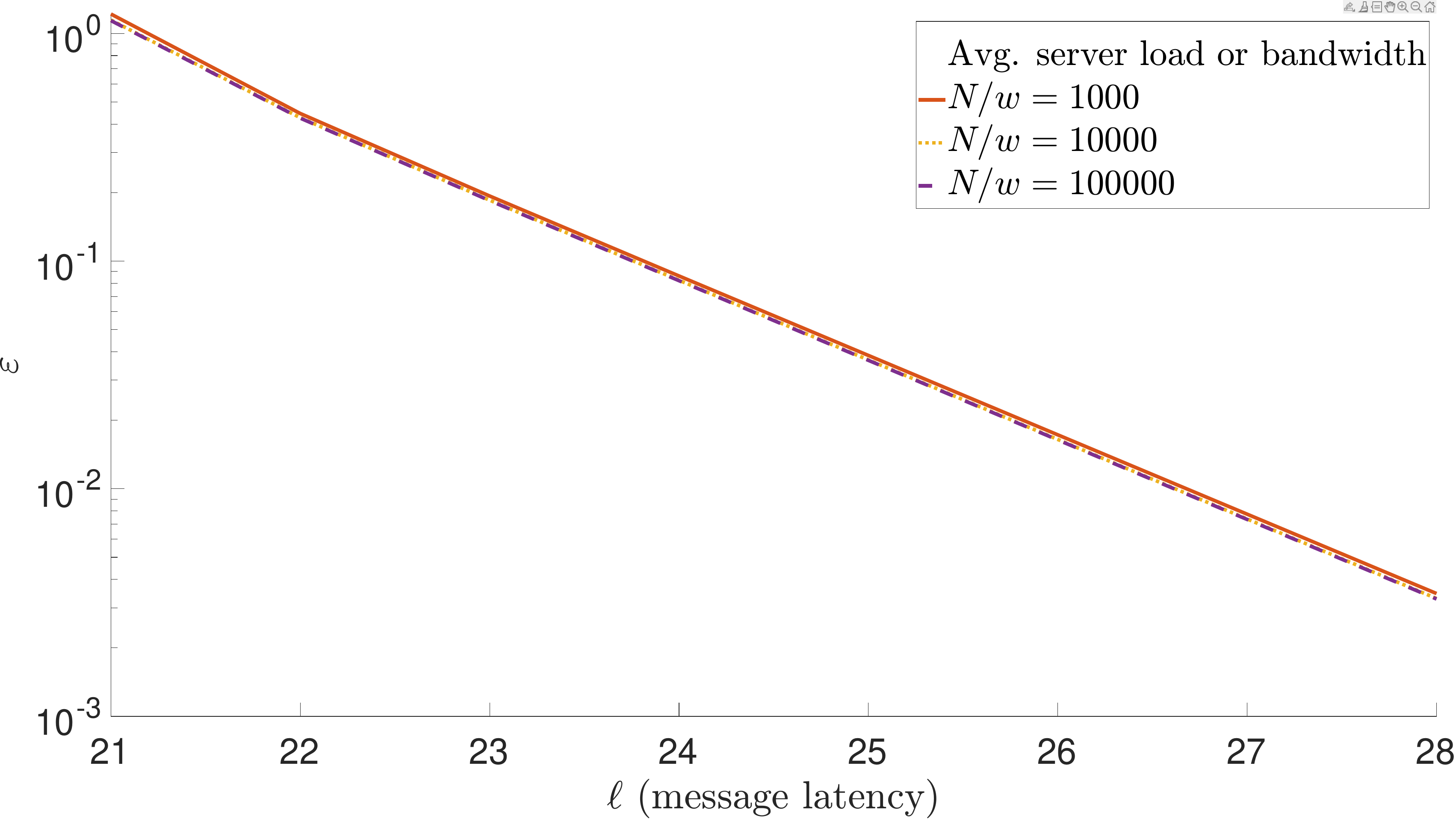}%
		\caption{$\alpha = 0.2$ and $\nmsgs = 10^6$.}
		\label{fig:M_20_8H}%
	\end{subfigure}%
	\begin{subfigure}{0.40\textwidth}
		\centering
		\includegraphics[width=\linewidth]{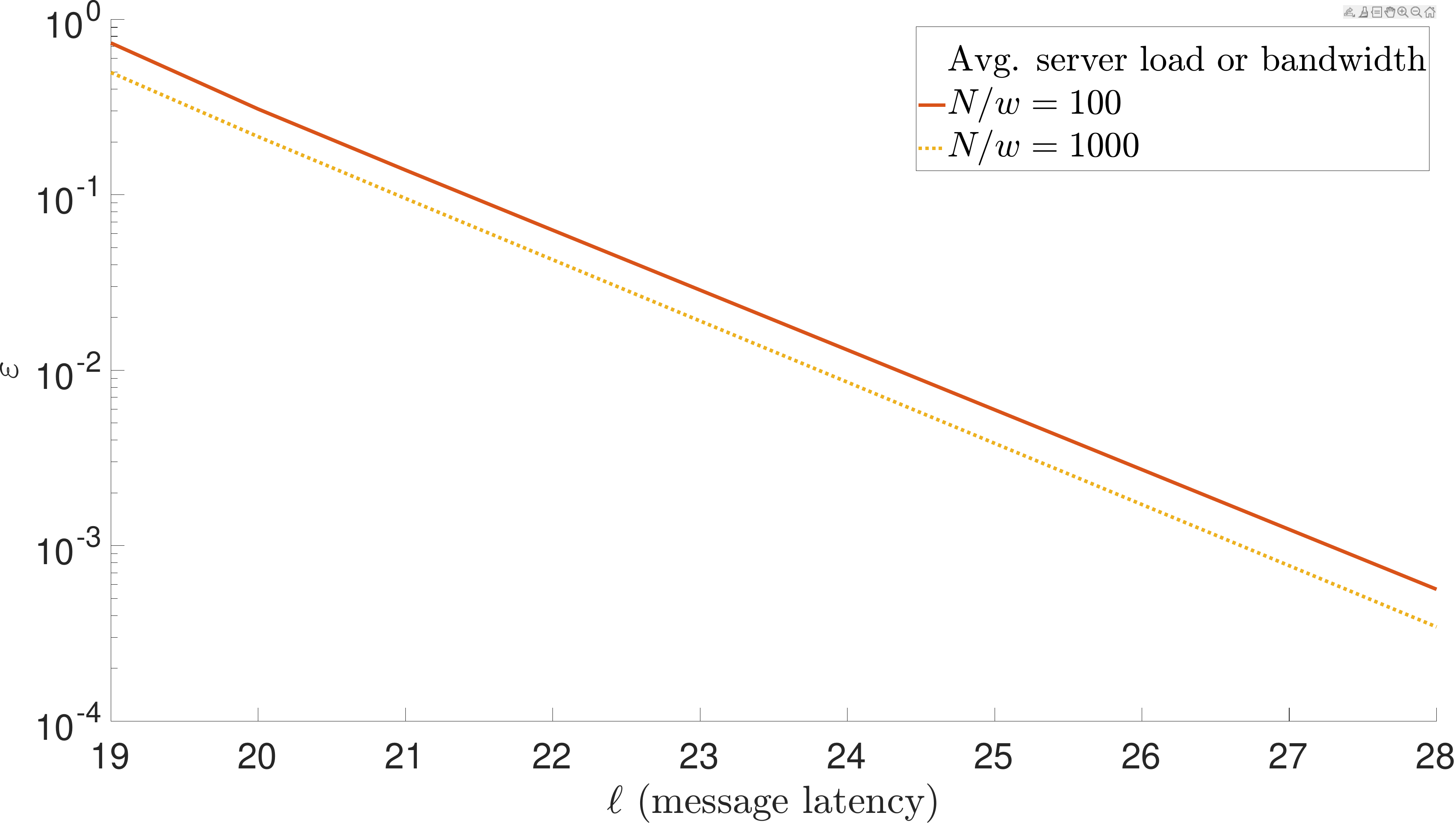}%
		\caption{$\alpha = 0.2$ and $\nmsgs = 10^4$.}
		\label{fig:M_20_8L}%
	\end{subfigure}%
	\centering%
	\caption{Provable minimum values of $\dpep$ (on a logarithmic scale) for varying $\pcompromise$, $\ndepth$ and $\frac{\nmsgs}{\nwidth}$. $\nmsgs$ is held fixed at either $10^6$ for the figures on the left or at $10^4$ for the figures on the right. $\dpde$ is fixed at $10^{-8}$.}%
	\label{fig:epsi_delta_delta8}
\end{figure*}

We now explain how the provable values of $\dpep$ and $\dpde$ change
in practice with the mixnet
parameters. 
\Cref{fig:epsi_delta_delta6,fig:epsi_delta_delta8} show the
exact values of $\dpep$ for two different values of $\dpde$ ($\dpde
\in \{10^{-6}, 10^{-8}\}$), 
several values of $\pcompromise$ (the server compromise
probability), $\ndepth$ (the length of the routing paths or message
latency), and $\nmsgs/\nwidth$ (average server load or average server
bandwidth). The values of $\dpep$ are calculated using
\cref{eqn:eps-del}. In each plot, we hold the number of messages
$\nmsgs$ fixed, and vary the server load $\nmsgs/\nwidth$ by varying
the number of servers $\nwidth$. The reason for this choice, as
opposed to holding $\nwidth$ fixed and varying $\nmsgs$, is that
$\nmsgs$ also sets the ideal against which we measure anonymity. In an
ideal mixnet, every posterior odds value $\postpr{\ndepth}{\mind}$ equals
$1/\nmsgs$, and $\dpep$ measures how far the actual posterior odds can
stray from this ideal. Increasing $\nmsgs$ therefore changes two
things at once: it lets each message mix with more messages, but it
also makes the ideal stricter. The second effect can make anonymity
appear to worsen as the load grows, even though messages are in fact
mixing better. Varying $\nwidth$ instead keeps the ideal the same in
each plot, so the individual curves isolate the effect of the server
load on mixing. We show two values of $\nmsgs$: $10^6$ (left column of
each figure) and $10^4$ (right column). The smaller $\nmsgs$ lets the
plots reach lower server loads, at which server compromise and server
load interact, as we discuss below.

\paragraph{Passive adversary ($\pcompromise = 0$)}
\Cref{fig:SH_6H}, \Cref{fig:SH_6L}, \Cref{fig:SH_8H} and \cref{fig:SH_8L} show $\dpep$ for $\dpde = 10^{-6}$
and $\dpde = 10^{-8}$ when no servers are compromised. Choosing
$\dpde = 10^{-6}$ yields excellent indistinguishability guarantees
(low $\dpep$) even for small values of $\ndepth$ and
$\frac{\nmsgs}{\nrouters}$. Specifically, for $\nmsgs = 10^6$, $\dpep$ ranges from $0.0631$ for $\ndepth = 5$ and $\frac{\nmsgs}{\nrouters} = 10^3$
all the way down to $1.06 \times 10^{-24}$ for $\ndepth = 12$ and
$\frac{\nmsgs}{\nrouters} = 10^5$.
For $\nmsgs = 10^4$, it ranges from $0.196$ for $\ndepth = 6$
and $\frac{\nmsgs}{\nrouters} = 100$ to $1.06 \times 10^{-13}$ for $\ndepth = 12$
and $\frac{\nmsgs}{\nrouters} = 10^3$.

Reducing $\dpde$ to $10^{-8}$ worsens $\dpep$, as expected due to the
approximation inherent to Chebyshev-based concentration bounds. However, we can still obtain strong indistinguishability bounds: for $\nmsgs = 10^6$, $\dpep$ ranges from $0.691$ for
$\ndepth = 5$ and $\frac{\nmsgs}{\nrouters} = 10^3$ to $1.06 \times 10^{-23}$ for
$\ndepth = 12$ and $\frac{\nmsgs}{\nrouters} = 10^5$. The
largest of these values is weak, since $e^{0.691} \approx 2$, but a
single additional round reduces it to $0.0199$ ($\ndepth = 6$,
$\frac{\nmsgs}{\nrouters} = 10^3$). So, only a
few rounds of mixing (5--7) suffice to provide strong anonymity
guarantees in the absence of server compromise.

\textit{Summary.}  For a passive adversary, our analysis yields very
strong provable anonymity even with short routing paths. The
quantitative trends confirm what is known from prior work: Improving
latency (reducing $\ndepth$) or improving per-router bandwidth
(reducing $\nmsgs/\nwidth$) increases the provable values of the
parameters $\dpep, \dpde$, thus reducing the degree of provable
anonymity. The effects are non-linear, clearly defined by
\cref{eqn:eps-del}.

\paragraph{Low server compromise rate ($\pcompromise = 1\%$)}
There is an immediate degradation (increase) in the provable parameter
$\dpep$ if we allow the adversary to compromise even a small fraction of
servers. As shown in \cref{fig:M_1_6H}, for $\nmsgs = 10^6$ and $\dpde
= 10^{-6}$, $\dpep$ ranges from
$0.283$ for $\frac{\nmsgs}{\nrouters} =
10^3$ and $\ndepth = 7$ to
$2.02 \times 10^{-12}$ for
$\frac{\nmsgs}{\nrouters} = 10^5$ and
$\ndepth = 18$ when $\pcompromise = 0.01$. For $\ndepth < 7$, we
cannot get any provable bounds~for $\nmsgs = 10^6$. The
reason for this is that with $\alpha > 0$, increasing $\nmsgs/\nwidth$
can reduce the variance of $\postpr{\ndepth}{\mind}$ only to a limited
extent: since $\beta \approx \nwidth/\nmsgs + \pcompromise$, the load
stops mattering once $\nmsgs/\nwidth$ is well above $1/\pcompromise =
100$. Indeed, the three curves for $\nmsgs = 10^6$, whose
  loads are all at least $10^3$, nearly coincide all the way from
$\ndepth = 7$ to $\ndepth = 18$. At $\ndepth = 18$, they differ by
  only about a factor of 2: $4.67 \times 10^{-12}$, $2.18 \times
10^{-12}$ and $2.02 \times 10^{-12}$ for $\frac{\nmsgs}{\nrouters} =
10^3, 10^4, 10^5$. In contrast, for $\nmsgs = 10^4$
  (\cref{fig:M_1_6L}), raising the load from $100 = 1/\pcompromise$ to
  $10^3$ still reduces $\dpep$ about 36-fold at $\ndepth = 12$, from
  $1.21 \times 10^{-5}$ to $3.34 \times 10^{-7}$. Thus, with
compromise, the only way to reduce the variance substantially
is to increase $\ndepth$.

Reducing $\dpde$ to $10^{-8}$ (\cref{fig:M_1_8L,fig:M_1_8H})
increases $\dpep$
  about tenfold at every point, since $\dpep$ scales with
  $1/\sqrt{\dpde}$ once it is small, and costs one more round of
  mixing: for $\nmsgs = 10^6$, provable bounds start at $\ndepth =
8$, and $\dpep$ ranges from $0.297$ ($\ndepth = 8$,
$\frac{\nmsgs}{\nrouters} = 10^3$) to $2.02 \times 10^{-11}$ ($\ndepth
= 18$, $\frac{\nmsgs}{\nrouters} = 10^5$).

\textit{Summary.} For a low server compromise rate (a weakly active
adversary), our analysis yields strong provable anonymity with
moderately long routing paths. The degree of anonymity continues to
improve if we allow for higher latency (higher $\ndepth$), but increasing
the server load (higher $\nmsgs/\nwidth$)
helps only up to $\frac{\nmsgs}{\nwidth} \approx
\frac{1}{\pcompromise}$, beyond which $\pcompromise$ dominates
$\beta$.

\paragraph{High server compromise rates ($\pcompromise = 10\%, 20\%$)}
\cref{fig:M_10_6H}, \cref{fig:M_10_6L}, \cref{fig:M_10_8H},
\cref{fig:M_10_8L}, and \cref{fig:M_20_6H}, \cref{fig:M_20_6L},
\cref{fig:M_20_8H} and \cref{fig:M_20_8L} show the graphs of
$\dpep$ with server compromise rates
$10\%$ and $20\%$, respectively. In these cases, $\beta$ is completely
dominated by $\pcompromise$, and the server load has almost no effect:
  for $\pcompromise = 10\%$, $\dpde = 10^{-6}$, $\nmsgs = 10^6$ and
  $\ndepth = 24$, $\dpep$ is $2.23 \times 10^{-6}$, $2.02 \times
  10^{-6}$ and $2.00 \times 10^{-6}$ for $\frac{\nmsgs}{\nrouters} =
  10^3, 10^4, 10^5$. Increasing $\ndepth$ continues to improve
anonymity strongly, though the minimum $\ndepth$ needed to get
provable guarantees is substantially higher: for
$\nmsgs = 10^6$, $\ndepth \geq 13$ for $\pcompromise = 10\%$ and
$\ndepth \geq 18$ for $\pcompromise = 20\%$ when $\dpde = 10^{-6}$,
and $\ndepth \geq 15$ and $\ndepth \geq 21$ when $\dpde = 10^{-8}$,
compared to $\ndepth \geq 7$ and $\ndepth \geq 8$ for
$\pcompromise = 1\%$.

\textit{Summary.}  Overall, we draw the following conclusion. With no
server compromise, the provable degree of anonymity can be improved by
increasing $\ndepth$ (latency) and by increasing
$\frac{\nmsgs}{\nwidth}$ (server load and bandwidth). With server
compromise, increasing $\ndepth$ continues to be a good strategy for
improving provable anonymity, but increasing server load beyond
$\frac{\nmsgs}{\nwidth} \approx \frac{1}{\pcompromise}$ offers little
additional privacy because $\pcompromise$ starts to dominate $\beta$.

\begin{table*}[tp]
  \centering
\begin{tabular}{l c c c c | c | c}
\toprule
\multicolumn{5}{c}{MixNet configuration~\cite{kate25tightrope}} & Empirical estimate~\cite{kate25tightrope} & Provable bound (ours) \\
\cmidrule(r){1-5} \cmidrule(lr){6-6} \cmidrule(l){7-7}
System & $\nmsgs$ & $\ndepth$ & $\nwidth$ & $\pcompromise$ & $\underline{\delta}(\infty) \le \tilde{\delta}(\infty) \le \overline{\delta}(\infty)$ & $\dpde$ at $\dpep = 10$ \\
\midrule
--      & 100 & 3 & 20 & 0.05 & 0.2 $\le$ 0.3 $\le$ 0.3       & $> 1$ (needs $\ndepth \geq 4$) \\
--      & 100 & 3 & 20 & 0.1  & 0.3 $\le$ 0.3 $\le$ 0.3       & $> 1$ (needs $\ndepth \geq 5$) \\
--      & 100 & 5 & 5  & 0.2  & 4e-3 $\le$ 6e-3 $\le$ 8e-3    & 0.13 \\
--      & 100 & 5 & 20 & 0.05 & 2e-3 $\le$ 3e-3 $\le$ 4e-3    & 0.13 \\
--      & 100 & 5 & 20 & 0.1  & 0.01 $\le$ 0.01 $\le$ 0.01    & 0.29 \\
\midrule
--      & 1M & 3 & 20   & 0.05 & 6e-3 $\le$ 7e-3 $\le$ 8e-3   & $> 1$ (needs $\ndepth \geq 5$) \\
--      & 1M & 3 & 20   & 0.1  & 0.02 $\le$ 0.03 $\le$ 0.03   & $> 1$ (needs $\ndepth \geq 7$) \\
--      & 1M & 3 & 1000 & 0.05 & 0.04 $\le$ 0.04 $\le$ 0.04   & $> 1$ (needs $\ndepth \geq 5$) \\
--      & 1M & 3 & 1000 & 0.1  & 0.08 $\le$ 0.09 $\le$ 0.09   & $> 1$ (needs $\ndepth \geq 7$) \\
\midrule
--      & 1M & 5 & 20   & 0.05 & 3e-7 $\le$ 2e-5 $\le$ 1e-4   & 0.63 \\
--      & 1M & 5 & 20   & 0.1  & 2e-4 $\le$ 4e-4 $\le$ 7e-4   & $> 1$ (needs $\ndepth \geq 7$) \\
--      & 1M & 5 & 20   & 0.2  & 5e-3 $\le$ 6e-3 $\le$ 9e-3   & $> 1$ (needs $\ndepth \geq 10$) \\
Atom    & 1M & 5 & 100  & 0.02 & 3e-36 $\le$ 5e-6 $\le$ 4e-4  & 6.6e-3 \\
--      & 1M & 5 & 1000 & 0.05 & 2e-5 $\le$ 2e-4 $\le$ 6e-4   & 0.69 \\
--      & 1M & 5 & 1000 & 0.1  & 9e-4 $\le$ 2e-3 $\le$ 3e-3   & $> 1$ (needs $\ndepth \geq 7$) \\
\midrule
--      & 1M & 10 & 20   & 0.05 & 0 $\le$ 0 $\le$ 3e-4       & 2.0e-7 \\
Atom    & 1M & 10 & 100  & 0.02 & 0 $\le$ 0 $\le$ 5e-4       & 2.1e-11 \\
Karaoke & 1M & 10 & 1000 & 0.2  & 0 $\le$ 0 $\le$ 1e-3       & 0.21 \\
Karaoke & 1M & 14 & 20   & 0.2  & 0 $\le$ 0 $\le$ 3e-4       & 3.3e-4 \\
Karaoke & 1M & 14 & 1000 & 0.2  & 0 $\le$ 0 $\le$ 3e-3       & 3.5e-4 \\
--      & 1M & 15 & 20   & 0.05 & 0 $\le$ 0 $\le$ 3e-4       & 6.1e-14 \\
Atom    & 1M & 15 & 100  & 0.02 & 0 $\le$ 0 $\le$ 6e-4       & 7.0e-20 \\
\bottomrule
\end{tabular}%
\caption{Our provable bounds on $\dpde$ at $\dpep = 10$, compared
  to the empirical estimates of $\dpde(\infty)$ of prior
  work~\cite{kate25tightrope}, for every configuration that prior work
  reports. Prior work treats every privacy loss of at least $10$ as
  $\infty$, so $\dpep = 10$ is the most conservative choice for our
  bound. The column $\pcompromise$ is the fraction $|C|/\nwidth$ of
  compromised servers in prior work. Where our bound exceeds $1$, it
  is vacuous; we give instead the minimum $\ndepth$ at which it falls
  below $1$.}
\label{tab:privacy_leakage_eps10}
\end{table*}

\paragraph{Comparison to empirical measurements of $\dpep, \dpde$}
Prior work~\cite{kate25tightrope} proposed a technique for
\emph{empirically} estimating $\dpde$, and a confidence interval
around it, for any given value of $\dpep$. Its setting is closely
analogous to ours: the adversary tries to tell which of two recipients
received a target message, which amounts to telling at which of two
positions the message left the mixnet. Those empirical estimates are
not mathematically provable bounds but it is, nonetheless, instructive
to compare them to our provable bounds. The comparison needs care,
however, because the two works measure leakage with different variants
of differential privacy (DP): ours uses probabilistic DP, while the
prior work uses a weaker variant, approximate DP. The two variants are
close for large $\dpep$ but differ significantly for $\dpep \approx 0$
and, in fact, our method (\cref{cor:indis_exit_msgs}) does not yield a
bounded $\dpde$ for $\dpep = 0$. Consequently, we only compare the two
works for large $\dpep$. (\cref{appendix:comparison-empirical}
explains this point further.)

\Cref{tab:privacy_leakage_eps10} shows --- for every mixnet
configuration reported by~\cite{kate25tightrope} (columns 1--5) ---
the empirical estimate of $\dpde$ for $\dpep = \infty$ and its
confidence interval as reported by~\cite{kate25tightrope} (column 6),
and the provable bound on $\dpde$ obtained by our method for $\dpep =
10$ (column 7). The symbols $\tilde{\delta}(\infty)$,
$\underline{\delta}(\infty)$ and $\overline{\delta}(\infty)$ denote
the empirically observed mean of $\delta$ and the lower and upper ends
of the confidence interval. The value $\dpep = \infty$ is used
by~\cite{kate25tightrope} as a proxy for $\dpep \geq 10$: every
privacy loss of at least $10$ is counted as $\infty$. Comparing our
bound at $\dpep = 10$ to these estimates is conservative for us in two
ways. First, our provable $\dpde$ decreases as $\dpep$ grows, so
$\dpep = 10$ yields the largest value of our bound in this range.
Second, at $\dpep = 10$ our bound is already
within $0.1\%$ of its limit $2(\nmsgs - 1)\beta^{\ndepth}$ as
$\dpep \to \infty$ (\cref{cor:indis_exit_msgs}) for all reported
configurations.

For short routing paths ($\ndepth \leq 5$), our provable values of
$\dpde$ are much worse (higher) than the empirical estimates, by
factors of about $20$ to $30{,}000$ where we get a bound at all. For many of
these configurations our bound exceeds $1$ and is therefore vacuous;
for those, \cref{tab:privacy_leakage_eps10} gives the minimum
$\ndepth$ at which our bound becomes meaningful. In particular, we get
no meaningful bound for any configuration with $\ndepth = 3$. This
difference is the price we pay for using a concentration bound to get
a \emph{provable} value of $\dpde$.

For longer routing paths ($\ndepth \geq 10$), which is where parallel
mixnets should operate if server compromise is a concern, the picture
changes. Prior work observed no distinguishing events at all in its
samples, so its empirical estimate of $\dpde$ is $0$, with an upper
confidence end between $3 \times 10^{-4}$ and $3 \times 10^{-3}$. Such
an estimate cannot distinguish events that are rare from events that
never happen. Our method, in contrast, yields a provable $\dpde$ for
every one of these configurations, and it is tiny in most of them:
from $2.0 \times 10^{-7}$ down to $7.0 \times 10^{-20}$. The exception
is Karaoke with $\pcompromise = 0.2$, where our bound is $0.21$ for
$\ndepth = 10$ and about $3.4 \times 10^{-4}$ for $\ndepth = 14$. The
reason is that for high $\ndepth$, $\postpr{\ndepth}{\mind}$ is very
narrowly concentrated around its mean, $1/\nmsgs$, unless
$\pcompromise$ is large, so our concentration bounds become very
tight.

To conclude, for parallel mixnets operating with low $\ndepth$, the
empirical method yields much smaller (but unprovable) estimates of
$\dpde$ than our method's provable bounds. As $\ndepth$ increases, our
method yields provable bounds that are very small for all but the most
heavily compromised configurations, while the empirical approach,
being based on sampling, observes no leakage at all.

\section{Our Stochastic Analysis and Proofs of Key Theorems}
\label{sec:analysis}

We now turn to our stochastic analysis of parallel mixnets,
specifically, an analysis of the properties of the random variable
$\postpr{\lind}{\mind}$ that eventually lead to the results of
\cref{sec:key-results}. We introduce some notation first.

Let $\roundinf_\lind$ be the distribution of the randomness that goes
into round $\lind$ of the parallel mixnet -- the selection of the
$\lind$-th round servers of all messages by the routing authority, the
shuffling permutations of the routers in round $\lind$, and the
permutation leading to $\idealarrs{\lind}$. 

Let $\partinf_\lind$ be the distribution of all the randomness used by
the abstract model of \cref{sec:mixnet-model} up to the construction
of $\idealarrs{\lind}$. Note that $\partinf_\lind$ is the
\emph{independent product} of
$\roundinf_1, \roundinf_2, \ldots, \roundinf_\lind$.

Finally, let $\partinf = \partinf_\ndepth$ be the distribution of all
the randomness used for a cohort. Then, a run $f$ is simply a sample
$f \sim \partinf$. We let $\obss_f$ denote the observation of the
adversary in the run $f$.

We let $\mrcmean{\lind}{\mind}$ denote the mean of the random variable
$\postpr{\lind}{\mind}$ across all possible runs. Further, we define
$\mrmeandiff{\lind}{\mind}$ as the deviation of
$\mrcmean{\lind}{\mind}$ from the ideal $\frac{1}{\nmsgs}$, and
$\mcvar{\lind}$ as the average squared sum of
$\mrmeandiff{\lind}{\mind}$ for all $\mind$. $\mcvar{\lind}$ is an
important random variable as it represents the squared error in the
deviation of the $\mrmeandiff{\lind}{\mind}$s from their ideal values.
\begin{eqnarray}
  \mrmeandiff{\lind}{\mind} & \defeq & \left(\postpr{\lind}{\mind} - \frac{1}{\nmsgs}\right) \\
  \mcvar{\lind} & \defeq & \frac{1}{\nmsgs} \cdot \sum_{\mind=1}^{\nmsgs} \mrmeandiff{\lind}{\mind}^2
\end{eqnarray}

We start by observing the following:
\begin{lemma}\label{lem:postpr-iden}
  $\postpr{\lind}{\mind}$ and $\postpr{\lind}{\mindB}$ are identically
  distributed for any $\mind, \mindB$.
\end{lemma}
\begin{proof}
  Immediate from the fact that $\postpr{\lind}{\mind}$ and
  $\postpr{\lind}{\mindB}$ are the same attribute of the elements at
  positions $\mind$ and $\mindB$ of an array, $\idealarrs{\lind}$,
  that has been shuffled uniformly at random.
\end{proof}

\subsection{Mean of $\postpr{\lind}{\mind}$}
We start by showing that the mean of $\postpr{\lind}{\mind}$, denoted
$\mrcmean{\lind}{\mind}$, is exactly $\frac{1}{N}$ for all
$\lind, \mind$. To prove this, we actually establish a stronger
property first, namely, that
$\sum_{\mind=1}^{\nmsgs} \postpr{\lind}{\mind} = 1$ for all $\lind$.

\begin{lemma}\label{lem:row-sum-const}
	For all $\lind$,
	$\sum_{\mind=1}^{\nmsgs} \postpr{\lind}{\mind} = 1$.
\end{lemma}
\begin{proof}
	We induct on $\lind$.
	
	\medskip
	\noindent \textit{Case} $\lind = 0$. Consider any run of the
	system. After the senders send their messages to $\rauth$, the
	latter shuffles the messages randomly with the random permutation
	$\mpermglob{0}$ to obtain $\idealarrs{0}$. Exactly one position in
	this array, say $\mind_0$, ends up with the adversary's message of
	interest, $\tgtmsg{}$. Then, $\postpr{0}{\mind_0} = 1$ and
	$\postpr{0}{\mind} = 0$ for $\mind \neq \mind_0$. Hence,
	$\sum_{\mind=1}^{\nmsgs} \postpr{\lind}{\mind} = 1$.
	
	\medskip
	\noindent \textit{Case} $\lind = \lindB + 1$. Let $\incidents{\lind}{\rind}$
	be the set of indices of $\idealarrs{\lind}$ whose messages go to
	router $\rind$ and let $\postpB_{\rind}$ be the posterior
	probability associated with every outgoing message of that router.
	Then,
	\[\postpB_{\rind} = \frac{\sum_{\mind \in \incidents{\lind}{\rind}}
		\postpr{\lind-1}{\mind}}{\left|\incidents{\lind}{\rind}\right|}
	\]
	Hence, the sum of posterior probabilities of all messages in
	$\idealarr{\lind}$ is:
	\[ \sum_{\rind=1}^\nwidth \left|\incidents{\lind}{\rind}\right| \cdot
	\frac{\sum_{\mind \in \incidents{\lind}{\rind}}
		\postpr{\lind-1}{\mind}}{\left|\incidents{\lind}{\rind}\right|}
	\]
	Since $\idealarrs{\lind}$ is an (adversary-visible) permutation of
	$\idealarr{\lind}$, the sum of posterior probabilities of elements
	of $\idealarrs{\lind}$ is the same as the expression above. Hence,
	\[
	\begin{tblr}{colspec={>{\displaystyle}ll>{\displaystyle}l},rowsep=3pt}
		\sum_{\mind=1}^{\nmsgs} \postpr{\lind}{\mind}
		& = &  \displaystyle\sum_{\rind=1}^\nwidth \left|\incidents{\lind}{\rind}\right| \cdot \frac{\sum_{\mind \in \incidents{\lind}{\rind}} \postpr{\lind-1}{\mind}}{\left|\incidents{\lind}{\rind}\right|} \\
		& = & \sum_{\rind=1}^\nwidth \sum_{\mind \in \incidents{\lind}{\rind}} \postpr{\lind-1}{\mind} \\
		& = & \sum_{\rind=1}^\nwidth \sum_{\mind \in \incidents{\lind}{\rind}} \postpr{\lindB}{\mind} \\
		& = & \sum_{\mind=1}^\nmsgs \postpr{\lindB}{\mind} \\
		& \stackrel{i.h.}{=} & 1
	\end{tblr}
	\]
	\mbox{}
\end{proof}

\begin{thm}\label{thm:prob-mean}
  For all $\lind \geq 0$ and $\mind$,
  $\expecd{\partinf_\lind}{\postpr{\lind}{\mind}} = \frac{1}{\nmsgs}$.
\end{thm}
\begin{proof}
	Start from \Cref{lem:row-sum-const}. Taking expectations on both
	sides and using the linearity of expectations, we get that
	$\sum_{\mind=1}^{\nmsgs}
	\expecd{\partinf_\lind}{\postpr{\lind}{\mind}} = 1$. Since the
	random variables $\postpr{\lind}{\mind}$ and
	$\postpr{\lind}{\mindB}$ are identically distributed
	(\Cref{lem:postpr-iden}), their expectations are equal, so
	$\nmsgs \cdot
	\left(\expecd{\partinf_\lind}{\postpr{\lind}{\mind}}\right) = 1$ for
	any $\mind$ and, hence,
	$\expecd{\partinf_\lind}{\postpr{\lind}{\mind}} = \frac{1}{\nmsgs}$,
	as required.
\end{proof}

\subsection{Variance of $\postpr{\lind}{\mind}$}

We now study the variance of $\postpr{\lind}{\mind}$. %
Our key insight is the following: Rather than analyze
$\postpr{\lind}{\mind}$ separately for each $\mind$, which yields very
loose bounds on the variance, we analyze the sum of squared errors,
$\mcvar{\lind}$, which combines the $\postpr{\lind}{\mind}$
together. This yields stronger, precise results.

The centerpiece of our technical work is the following theorem, which
relates the expected value of $\mcvar{\lind}$ to the expected value of
$\mcvar{\lind-1}$ \emph{precisely}. Specifically, the theorem shows
that this expected value falls by a constant factor ($\beta$) in each
round of forwarding. The proof is nontrivial, but uses elementary
methods. It is also instructive, and we encourage the enthusiastic
reader to read the proof.

\begin{thm}\label{thm:prob-var-exp}
	For any $\lind \in \{1,\ldots,\ndepth\}$,
	\[
	\expecd{\partinf_\lind}{\mcvar{\lind}} =
	\beta \cdot
	\expecd{\partinf_{\lind-1}}{\mcvar{\lind-1}}
	\]
	where $\beta$ is 
	\[
	\beta = \contfact{\nmsgs}{\nwidth} \cdot \left(1 - \alpha\right) + \alpha
	\]
	and $\contfact{\nmsgs}{\nwidth}$ is the following constant
	independent of $\lind$:
	\[
	\contfact{\nmsgs}{\nwidth} = \frac{\nwidth-1}{\nmsgs-1}\left[ 1 - \left(1 - \frac{1}{\nwidth} \right)^{\nmsgs-1} \right] < \frac{\nwidth}{\nmsgs}
	\]
\end{thm}
\begin{proof} 
	Let $f_{\lind-1}$ denote an element of the sample space of
	$\partinf_{\lind-1}$.
	By the law of total expectations:
	\[
	\begin{array}{@{}l@{\;}l@{\;}l@{}}
		\expecd{\partinf_\lind}{\mcvar{\lind}} & = &
		\expecd{f_{\lind-1} \sim \partinf_{\lind-1}}{\expecd{\roundinf_{\lind}}{\pcond{\mcvar{\lind}}{\left(\partinf_{\lind-1} = f_{\lind-1}\right)}}}
	\end{array}
	\]
	We first calculate
	$\expecd{\roundinf_{\lind}}{\pcond{\mcvar{\lind}}{\left(\partinf_{\lind-1}
			= f_{\lind-1}\right)}} =
	\expecd{\roundinf_{\lind}}{\pcond{\mcvar{\lind}}{E_{\lind-1}}}$ for
	a fixed $f_{\lind-1}$, where $E_{\lind-1}$ is the event
	$\partinf_{\lind-1} = f_{\lind-1}$. We have\\
	\begin{tblr}{colspec={Q[l,$]Q[l,$]Q[l,$$]}}
		&   & \expecd{\roundinf_{\lind}}{\pcond{\mcvar{\lind}}{E_{\lind-1}}}\\
		& = & \expecd{\roundinf_{\lind}}{\pcond{\frac{1}{\nmsgs} \cdot \sum_{\mind=1}^{\nmsgs} \mrmeandiff{\lind}{\mind}^2}{E_{\lind-1}}}\\
	\end{tblr}\\
	Let
	$\exigents{\lind}{\rind}$ be the set of message indexes that arrive
	in $\idealarrs{\lind}$ from router
	$\mrouter{\lind}{\rind}$ (via
	$\idealarr{\lind}$). We split the sum along $\mrouter{\lind}{\rind}$.
        \[
	\begin{tblr}{colspec={Q[l,$]Q[l,$]Q[l,$$]}}
		& = & \expecd{\roundinf_{\lind}}{\pcond{\frac{1}{\nmsgs} \cdot \sum_{\rind=1}^{\nrouters}\sum_{\mindB\in \exigents{\lind}{\rind}} \mrmeandiff{\lind}{\mindB}^2} {E_{\lind-1}}}
	\end{tblr}
        \]
	Let $\incidents{\lind}{\rind}$ be the set of indices in
	$\idealarrs{\lindB}$, where $\lindB = \lind-1$, which are sent to
	$\mrouter{\lind}{\rind}$. Now, we separate the inner term of the expected value between compromised and uncompromised routers.
	Using linearity of expectations and using the averaging on honest routers we rewrite
	the above expression in terms of round $\incidents{\lind}{\rind}$.
	A compromised router does no averaging: the adversary
	knows its permutation, so the router maps the deviations of its
	incoming messages one-to-one to those of its outgoing messages.
        \[
        \begin{tblr}{colspec={Q[l,$]Q[l,$]Q[l,$$]}}
	  & = & \expecd{\roundinf_{\lind}}{\pcond{\frac{1}{\nmsgs} \cdot \sum_{\rind=1}^{\nrouters}\nrmsgs{\lind}{\rind} \cdot F \left(N_{i,h}\right)}{E_{\lind-1}}} \\      
	\end{tblr}
        \]
	\medskip
	\noindent
	Where $F\left(N_{i,h}\right)$ is defined as follows ($\mathbb{I}(x)$ is the indicator function that maps false to $0$ and true to $1$)
	\begin{align*}
		F\left(N_{i,h}\right) =~ & \mathbb{I}\left[\router_h \notin \alpha \right] \cdot \left(\frac{\sum_{\mindB \in \incidents{\lind}{\rind}} \mrmeandiff{\lindB}{\mindB}}{\nrmsgs{\lind}{\rind}}\right)^2 + \\
		& \mathbb{I}\left[\router_h \in \alpha \right] \cdot
		\frac{\sum_{\mindB \in \incidents{\lind}{\rind}} \mrmeandiff{\lindB}{\mindB}^2}{\nrmsgs{\lind}{\rind}}
	\end{align*}
	Next, we come to the key step in the proof. Let
	$\sentto{\lindB}{\mind}$ be the index of the router to which
	$\idealarrs{\lindB+1}[\mind]$ is routed in stage
	$\lindB+1$. Further, let
	$\samerouter{\lindB}{\mindC,\mindD}$ be a predicate which is true if
	$\idealarrs{\lindB}[\mindC]$ and
	$\idealarrs{\lindB}[\mindD]$ land on the same router in stage
	$\lindB+1$, i.e., if
	$\sentto{\lind}{\mindC}=\sentto{\lind}{\mindD}$, and let
	$\indfn{\samerouter{\lindB}{\mindC,\mindD}}$ be this predicate's
	indicator function, which is $1$ if the predicate is true, else
	$0$. Then, it is not hard to see that the above expression can be
	rewritten as:
        \[
	\begin{tblr}{colspec={Q[l,$]Q[l,$]Q[l,$$]}}
		& = & \mathbb{E}_{\roundinf_{\lind}}\Biggl[\frac{1}{N}\sum_{\mindC,\mindD \in [\nmsgs]} \mrmeandiff{\lindB}{\mindC} \mrmeandiff{\lindB}{\mindD} \frac{\indfn{\samerouter{\lindB}{\mindC,\mindD}}}{\nrmsgs{\lind}{\sentto{\lindB}{\mindC}}} \\
		& & \hspace*{2em}{} \cdot \mathbb{I}\left[\router_h \notin \alpha \right] \;\Bigg|\; E_{\lind-1}\Biggr] + \\
		& + &  \\
		& & \expecd{\roundinf_{\lind}}{\pcond{\frac{1}{N}\sum_{\mindC \in [\nmsgs]} \mrmeandiff{\lindB}{\mindC}^2 \cdot \mathbb{I}\left[\router_h \in \alpha \right]}{E_{\lind-1}}}
	\end{tblr}
        \]
	We consider the first part of the above equation now, the conditioning on $E_{\lind-1}$ makes $\mrmeandiff{\lindB}{\mindC}$ and $\mrmeandiff{\lindB}{\mindD}$ \emph{constants}, so we can pull them out of the expectation to obtain
        \[
	\begin{tblr}{colspec={Q[l,$]Q[l,$]Q[l,$$]}}
		& = & \frac{1}{N}\sum_{\mindC,\mindD \in [\nmsgs]} \mrmeandiff{\lindB}{\mindC} \mrmeandiff{\lindB}{\mindD} \\
		& & \hspace*{1em}{} \cdot \expecd{\roundinf_{\lind}}{\pcond{\frac{\indfn{\samerouter{\lindB}{\mindC,\mindD}}}{\nrmsgs{\lind}{\sentto{\lindB}{\mindC}}} \cdot \mathbb{I}\left[\router_h \notin \alpha \right] }{E_{\lind-1}}}
	\end{tblr}
        \]
	Since
	$\frac{\indfn{\samerouter{\lindB}{\mindC,\mindD}}}{\nrmsgs{\lind}{\sentto{\lindB}{\mindC}}}$
	is independent of
	$E_{\lind-1}$, we can get rid of the conditioning. Splitting the sum
	on whether $\mindC = \mindD$ or not,
        \[
	\begin{tblr}{colspec={Q[l,$]Q[l,$]Q[l,$$]}}\\
		& = & \frac{1}{N}\sum_{\mindC \in [\nmsgs]} \mrmeandiff{\lindB}{\mindC}^2 \cdot \expecd{\roundinf_{\lind}}{\frac{1}{\nrmsgs{\lind}{\sentto{\lindB}{\mindC}}} \cdot \mathbb{I}\left[\router_h \notin \alpha \right] } + \\
		& & \frac{1}{N}\sum_{\mindC \neq \mindD \in [\nmsgs]}
		\mrmeandiff{\lindB}{\mindC} \mrmeandiff{\lindB}{\mindD} \\
		& & \hspace*{1em}{} \cdot
		\expecd{\roundinf_{\lind}}{\frac{\indfn{\samerouter{\lindB}{\mindC,\mindD}}}{\nrmsgs{\lind}{\sentto{\lindB}{\mindC}}} \cdot \mathbb{I}\left[\router_h \notin \alpha \right] }
	\end{tblr}
        \]
	Now, since the number of messages that a router receives in a round is independent of whether the router is compromised or not, we can rewrite the above equation as
        \[
	\begin{tblr}{colspec={Q[l,$]Q[l,$$]}}\\
	  = & \frac{1 - \alpha}{N}\sum_{\mindC \in [\nmsgs]} \mrmeandiff{\lindB}{\mindC}^2 \cdot \expecd{\roundinf_{\lind}}{\frac{1}{\nrmsgs{\lind}{\sentto{\lindB}{\mindC}}}} + \\
	  & \frac{1 - \alpha}{N}\sum_{\mindC \neq \mindD \in [\nmsgs]}
	  \mrmeandiff{\lindB}{\mindC} \mrmeandiff{\lindB}{\mindD} \cdot
	  \expecd{\roundinf_{\lind}}{\frac{\indfn{\samerouter{\lindB}{\mindC,\mindD}}}{\nrmsgs{\lind}{\sentto{\lindB}{\mindC}}}}
	\end{tblr}
        \]
	Now,
	$\expecd{\roundinf_{\lind}}{\frac{1}{\nrmsgs{\lindB}{\sentto{\lindB}{\mindC}}}}$
	is the expected number of messages on the router
	$\sentto{\lindB}{\mindC}$, \emph{given that it has at least one
		message}, namely, $\idealarrs{\lindB}[\mindC]$. Due to the random
	assignment of messages to routers, this expectation is exactly
	$\expec{1/(1+X)}$, where
	$X \sample \mdistbinom{\nmsgs-1}{1/\nrouters}$, which is easily
	calculated (see \Cref{thm:inverse-binom-expec-inst}(1)).
	\[
	\expecd{\roundinf_{\lind}}{\frac{1}{\nrmsgs{\lind}{\sentto{\lindB}{\mindC}}}}
	=
	\frac{1-(1-p)^\nmsgs}{\nmsgs p}
	\]
	where $p = 1/\nwidth$.
	
	On the other hand,
	$\expecd{\roundinf_{\lind}}{\frac{\indfn{\samerouter{\lindB}{\mindC,\mindD}}}{\nrmsgs{\lind}{\sentto{\lindB}{\mindC}}}}$
	is the product of the probability that both
	$\idealarrs{\lindB}[\mindC]$ and $\idealarrs{\lindB}[\mindD]$ land
	on the same router, which is exactly $1/\nrouters$, and the expected
	value of the reciprocal of the number of messages on that router,
	\emph{given that it has at least two messages}, namely,
	$\idealarrs{\lindB}[\mindC]$ and $\idealarrs{\lindB}[\mindD]$. The
	latter value is $\expec{1/(2+X)}$, where
	$X \sample \mdistbinom{\nmsgs-2}{1/\nrouters}$. Using
	\Cref{thm:inverse-binom-expec-inst}(2), we get:
	\[
	\expecd{\roundinf_{\lind}}{\frac{\indfn{\samerouter{\lindB}{\mindC,\mindD}}}{\nrmsgs{\lind}{\sentto{\lindB}{\mindC}}}}
	=
	\frac{1}{\nrouters} \cdot \frac{\nmsgs p -1 + (1-p)^\nmsgs}{\nmsgs(\nmsgs-1)p^2} 
	\]
	Combining everything together, we get
        \[
        \begin{tblr}{colspec={Q[l,$]Q[l,$$]}}
		& \expecd{\roundinf_{\lind}}{\pcond{\mcvar{\lind}}{E_{\lind-1}}} \\
		= & \frac{1-(1-p)^\nmsgs}{\nmsgs p} \cdot \frac{1 - \alpha}{\nmsgs}
		\sum_{\mindC \in
			[\nmsgs]} \mrmeandiff{\lindB}{\mindC}^2 \\
		& \mathop{+} \frac{1 - \alpha}{\nrouters} \cdot \frac{\nmsgs p -1 +
			(1-p)^\nmsgs}{\nmsgs(\nmsgs-1)p^2} \\
		& \hspace*{1em}{} \cdot
		\frac{1}{\nmsgs}\sum_{\mindC \neq \mindD \in [\nmsgs]}
		\mrmeandiff{\lindB}{\mindC} \mrmeandiff{\lindB}{\mindD}
	\end{tblr}
        \]
	Next, we consider 
	\[
		\expecd{\roundinf_{\lind}}{\pcond{\frac{1}{N}\sum_{\mindC \in [\nmsgs]} \mrmeandiff{\lindB}{\mindC}^2 \cdot \mathbb{I}\left[\router_h \in \alpha \right]}{E_{\lind-1}}},
	\]
	which can be trivially reduced to
	\begin{equation*}
		\frac{\alpha}{N} \cdot \sum_{\mindC \in [\nmsgs]} \mrmeandiff{\lindB}{\mindC}^2
	\end{equation*}
	because each router of round $\lind$ is compromised with
	probability $\alpha$, independently of $E_{\lind-1}$ (the servers of
	round $\lind$ form a layer of their own;
	see \cref{sec:threat-model}).
	Returning to the top-level proof,
        \[
        \begin{tblr}{colspec={Q[l,$$]}}
		\expecd{\partinf_\lind}{\mcvar{\lind}} \\
		= \expecd{f_{\lind-1} \sim \partinf_{\lind-1}}{\expecd{\roundinf_{\lind}}{\pcond{\mcvar{\lind}}{\left(\partinf_{\lind-1} = f_{\lind-1}\right)}}} \\
		= \frac{1-(1-p)^\nmsgs}{\nmsgs p}
		\cdot \left(1 - \alpha\right) \cdot \expecd{\partinf_{\lind-1}}{\frac{1}{\nmsgs} \sum_{\mindC \in
				[\nmsgs]} \mrmeandiff{\lindB}{\mindC}^2} \\
		\mathop{+} \frac{\nmsgs p -1 +
			(1-p)^\nmsgs}{\nrouters \nmsgs(\nmsgs-1)p^2} \cdot \left(1 - \alpha\right) \\
		\hspace*{1em}{} \cdot
		\expecd{\partinf_{\lind-1}}{\frac{1}{\nmsgs}\sum_{\mindC \neq \mindD \in [\nmsgs]}
			\mrmeandiff{\lindB}{\mindC} \mrmeandiff{\lindB}{\mindD}} \\
		\mathop{+} \alpha \cdot  \\
		\hspace*{1em} \expecd{\partinf_{\lind-1}}{\frac{1}{\nmsgs}\sum_{\mindC \in [\nmsgs]} \mrmeandiff{\lindB}{\mindC}^2}
	\end{tblr}
        \]
	Observe that
	$\expecd{\partinf_{\lind-1}}{\frac{1}{\nmsgs} \sum_{\mindC \in
			[\nmsgs]} \mrmeandiff{\lindB}{\mindC}^2}$, which
	appears in the first and the last terms, is exactly
	$\expecd{\partinf_{\lind-1}}{\mcvar{\lind-1}}$ (since
	$\lindB = \lind-1$).
	Further,
	$\expecd{\partinf_{\lind-1}}{\frac{1}{\nmsgs}\sum_{\mindC \neq
			\mindD \in [\nmsgs]} \mrmeandiff{\lindB}{\mindC}
		\mrmeandiff{\lindB}{\mindD}}$ $ = \\ $
	$\expecd{\partinf_{\lind-1}}{\frac{1}{\nmsgs}\left(\left(\sum_{\mindC
			\in [\nmsgs]} \mrmeandiff{\lindB}{\mindC}\right)^2 -
		\sum_{\mindC \in [\nmsgs]}
		\mrmeandiff{\lindB}{\mindC}^2\right)}$. Since
	$\sum_{\mindC \in [\nmsgs]} \mrmeandiff{\lindB}{\mindC} = 0$ by
	\Cref{lem:row-sum-const}, we get
	$\expecd{\partinf_{\lind-1}}{\frac{1}{\nmsgs}\sum_{\mindC \neq
			\mindD \in [\nmsgs]} \mrmeandiff{\lindB}{\mindC}
		\mrmeandiff{\lindB}{\mindD}} =
	-\expecd{\partinf_{\lind-1}}{\mcvar{\lind-1}}$.
	\medskip
	\noindent Hence,
        \[
	\begin{tblr}{colspec={Q[l,$]Q[l,$$]}}
		& \expecd{\partinf_\lind}{\mcvar{\lind}} \\
		= & \left( \frac{1-(1-p)^\nmsgs}{\nmsgs p} - \frac{\nmsgs p -1 +
			(1-p)^\nmsgs}{\nrouters \nmsgs(\nmsgs-1)p^2} \right) \\
		&  \cdot \left(1 - \alpha \right) \cdot \expecd{\partinf_{\lind-1}}{\mcvar{\lind-1}} \\
		& + \alpha \cdot \expecd{\partinf_{\lind-1}}{\mcvar{\lind-1}}
	\end{tblr}
      \]
      
	Using $p = 1/\nrouters$ and simplifying, we get,
        \[
        \begin{tblr}{colspec={Q[l,$]Q[l,$$]}}
	& \expecd{\partinf_\lind}{\mcvar{\lind}} \\
         = & \left(1 - \alpha \right) \cdot \frac{\nrouters-1}{\nmsgs-1}
	\left[1-\left(1-\frac{1}{w}\right)^{\nmsgs-1}\right] \\
	 & \hspace*{1em}{} \cdot
	\expecd{\partinf_{\lind-1}}{\mcvar{\lind-1}} \\
	 & \mbox{} + \alpha \cdot \expecd{\partinf_{\lind-1}}{\mcvar{\lind-1}} \\
          = & \left[ \left(1 - \alpha \right) \contfact{\nmsgs}{\nrouters} + \alpha \right] \cdot \expecd{\partinf_{\lind-1}}{\mcvar{\lind-1}} \\
          = & \beta \cdot
        	\expecd{\partinf_{\lind-1}}{\mcvar{\lind-1}}
	\end{tblr}
      \]
    \end{proof}

Using straightforward induction on $\lind$, we can derive the
following ``$\lind$-step'' relation between $\expec{\mcvar{\lind}}$
and $\expec{\mcvar{0}}$.

\begin{cor}\label{cor:prob-var-exp:2} The following holds for all
	$\lind \in \{1, \ldots, \ndepth\}$.
	\begin{enumerate}
		\item
		$\expec{\mcvar{0}} = \frac{1}{\nmsgs} \cdot \left( 1 -
		\frac{1}{\nmsgs} \right)$
		\item
		$\expecd{\partinf_\lind}{\mcvar{\lind}} =
		\beta^{\lind} \cdot
		\frac{1}{\nmsgs} \cdot \left( 1 - \frac{1}{\nmsgs}\right)$
	\end{enumerate} 
\end{cor}
\begin{proof} (1) In any run of the system, exactly one $\postpr{0}{j}$
	is $1$ and the remaining $\postpr{0}{j}$s are $0$. Hence,
	$\mcvar{0}$ $\defeq$ $\frac{1}{\nmsgs}\sum_{\mind=1}^{\nmsgs}
	\left(\postpr{0}{\mind} - \frac{1}{\nmsgs}\right)^2$ $=$\\ $\frac{1}{\nmsgs}
	\left[ \left(1-\frac{1}{\nmsgs}\right)^2 + (\nmsgs-1) \left( \frac{1}{\nmsgs}
	\right)^2 \right]$ $=$ $\frac{1}{\nmsgs} \cdot \left( 1 -
	\frac{1}{\nmsgs} \right)$.
	
	Thus, $\mcvar{0}$ is the constant
	$\frac{1}{\nmsgs} \cdot \left( 1 - \frac{1}{\nmsgs} \right)$. So, its
	expectation is also equal to the same value.
	
	\medskip
	\noindent (2) Immediate by induction on $\lind$, using
	\Cref{thm:prob-var-exp} for the induction step, and (1) for the base
	case $i=0$.

\end{proof}

\paragraph{Variance of $\postpr{\lind}{\mind}$}
Remarkably, \Cref{cor:prob-var-exp:2} allows us to calculate the
variance of $\postpr{\lind}{\mind}$, once again exploiting the
identical distribution of $\postpr{\lind}{\mind}$ for different
$\mind$ (\cref{lem:postpr-iden}).

\begin{manualtheorem}{\ref{thm:prob-var}}[Slightly reworded]
	For any $\lind, \mind$,
	\[
	\varnd{\partinf_{\lind}}{\postpr{\lind}{\mind}} =
	\expecd{\partinf_{\lind}}{\mcvar{\lind}} =
	\beta^{\lind} \cdot
	\frac{1}{\nmsgs} \cdot \left( 1 - \frac{1}{\nmsgs}\right)
	\]
\end{manualtheorem}
\begin{proof}
	The key idea is to consider the sum of variances
	$s = \sum_{\mind=1}^{\nmsgs}
	\varnd{\partinf_{\lind}}{\postpr{\lind}{\mind}}$. Since the random
	variables $\{\postpr{\lind}{\mind}\}_{\mind=1}^{\nmsgs}$ are
	identically distributed,
	$s = \nmsgs \cdot
	\varnd{\partinf_{\lind}}{\postpr{\lind}{\mind}}$. On the other hand:
	\[
	\begin{tblr}{colspec={>{\displaystyle}ll>{\displaystyle}lr}}
		s & = & \sum_{\mind=1}^{\nmsgs} \varnd{\partinf_{\lind}}{\postpr{\lind}{\mind}}\\
		& = & \sum_{\mind=1}^{\nmsgs} \expecd{\partinf_{\lind}}{\left( \postpr{\lind}{\mind} - \frac{1}{\nmsgs}\right)^2} \\
		& = & \expecd{\partinf_{\lind}}{\sum_{\mind=1}^{\nmsgs} \left( \postpr{\lind}{\mind} - \frac{1}{\nmsgs}\right)^2} \\
		& \defeq & \expecd{\partinf_{\lind}}{\nmsgs \cdot \mcvar{\lind}}  \\
		& = & \nmsgs \cdot \expecd{\partinf_{\lind}}{\mcvar{\lind}}
	\end{tblr}
	\]
	
	Therefore,
	$\nmsgs \cdot \varnd{\partinf_{\lind}}{\postpr{\lind}{\mind}} = s =
	\nmsgs \cdot \expecd{\partinf_{\lind}}{\mcvar{\lind}}$ and, hence,
	$\varnd{\partinf_{\lind}}{\postpr{\lind}{\mind}} =
	\expecd{\partinf_{\lind}}{\mcvar{\lind}}$. The required result
	follows from \Cref{cor:prob-var-exp:2}(2).
\end{proof}

\subsection{Proofs of Key Theorems}

\cref{cor:tail-bound-expec1} is an easy consequence of
\cref{thm:prob-var}, using a concentration bound, specifically,
Chebyshev's inequality. Note that in the corollary here, we have taken
the probabilities over $\partinf_\lind$, but this is equivalent to
taking probabilities over $\partinf$ because $\postpr{\lind}{\mind}$
only depends on the random choices included in $\partinf_\lind$.

\begin{manualcorollary}{\ref{cor:tail-bound-expec1}}[Restated]
	For any $0 < a < \frac{1}{\nmsgs}$,
	\[ \begin{aligned}
	\prd{\partinf_\lind}{ \; \left| \, \postpr{\lind}{\mind}-\frac{1}{\nmsgs} \, \right| \; \geq a}
	&\leq
	\beta^{\lind} \frac{1}{\nmsgs a^2} \left(1 - \frac{1}{\nmsgs}\right) \\
	&<
	\frac{1}{\nmsgs a^2} \beta^{\lind}
	\end{aligned} \]
\end{manualcorollary}
\begin{proof}
	From \Cref{thm:prob-mean} and \Cref{thm:prob-var}, the mean and
	variance of $\postpr{\lind}{\mind}$ are $\frac{1}{\nmsgs}$ and
	$\beta^{\lind} \cdot
	\frac{1}{\nmsgs} \cdot \left( 1 - \frac{1}{\nmsgs}\right)$,
	respectively. By Chebyshev's inequality,
	\[
	\begin{tblr}{colspec={@{}>{\displaystyle}l@{\,}l@{\,}>{\displaystyle}l}}
		\prd{\partinf_\lind}{\; \left| \, \postpr{\lind}{\mind}-\frac{1}{\nmsgs} \, \right| \; \geq a} & \leq &
		\frac{1}{a^2}\beta^{\lind} \cdot \frac{1}{\nmsgs} \cdot \left( 1 - \frac{1}{\nmsgs}\right) \\
		& < & \frac{1}{\nmsgs a^2} \left[ \left(1 - \alpha \right) \cdot \frac{\nwidth}{\nmsgs} + \alpha\right]^{\lind}
	\end{tblr}
	\]
\end{proof}

\cref{cor:indis_exit_msgs} follows from \cref{cor:tail-bound-expec1}
but it needs a deeper analysis that also accounts for
\emph{covariances} between $\postpr{\lind}{\mind}$s for different
$\mind$. We defer the proof to \Cref{app:analysis-honest}.

\section{Related work}
\label{sec:related-work}

Mixnets have been analyzed widely using a variety of techniques. We
compare here to the most closely related work on the analysis of
parallel mixnets.

In terms of end-goals, the most closely related work to ours is the
proof of differential privacy for a parallel mixnet design in
Karaoke~\cite{lazar18karaoke} (and related
systems~\cite{vuvuzela15mixnet,tyagi16stadium,barman22groove}). Karaoke
uses a highly innovative relational indistinguishability argument to
establish differential privacy for message routing. They calculate the
probability that any two messages satisfy a specific condition which
makes them indistinguishable: Either the messages are co-resident on
the same server in some round, or there are cross messages that cross
the trajectories of both the messages in a specific way. In order to
establish the existence of these cross messages, Karaoke requires
chaff messages between servers. Our result, on the other hand, is
targeted at a vanilla parallel mixnet without chaff
messages. Further, later work~\cite{DBLP:conf/csfw/DasMMK24} found a
fundamental flaw in Karaoke's proof argument. That later work proposed
and proved anonymity for a different mixnet design based on funnelling
of messages, but that design uses a specific routing scheme and does
not cover the class of mixnets we do.

In terms of analysis technique, our work is closest to the work of
Goodrich and Mitzenmacher (GM)~\cite{goodrich12cardspub}, who also
observed that the sum of the square deviations from the mean for
$\postpr{\lind}{\mind}$ (quantity $\mcvar{\lind}$ from
\cref{sec:analysis}) can be bounded as a function of the mixnet
rounds. However, their analysis is limited to the case where messages
are perfectly evenly distributed among the servers in each routing
round. An unpublished extension presents the general case, but the
expression for $\mcvar{\lind}$ is open, and complicated. Our analysis
yields a simple closed-form expression and, importantly, we use the
analysis to show a strong end-to-end anonymity property, and explore
its consequences in practice, which GM do not do.

The underlying structure of our model, specifically, the treatment of
posterior odds as random variables, mirrors the underlying structure
of the empirical analysis for parallel mixnets proposed in prior
work~\cite{kate25tightrope}. We presented a quantitative comparison to
this work in \cref{sec:evaluation}. Broadly speaking, we perform a
theoretical analysis of the posterior odds, while the prior work
proposes an experimental technique to measure the odds in simulations
and build confidence intervals around them.

Further back, Das \emph{et al.}~\cite{DBLP:conf/sp/Das0MK18} showed
that there is a fundamental upper bound on the degree of anonymity a
mixnet can offer, as a function of the tolerated latency and
per-server bandwidth.  Compared to our work, this work has the opposite goal:
We show a lower bound on anonymity in order to argue that parallel
mixnets have provably useful operating points, while Das \emph{et al.}
establish limits on what mixnets can do.

There is also a large body of work on a different measure of
anonymity, namely, entropy of
$\postpr{\lind}{\mind}$~\cite{mixim21mixnetsim,rahimi25mocha,nym21whitepaper}. Entropy
is an aggregate (averaged) anonymity property that does not lead to
guarantees for directed attacks (unlike our work and the work listed
above). Moreover, most work on entropy aims to measure it empirically,
not prove bounds on it.

Finally, there is work on mixnets that follow a fundamentally
different design from the parallel mixnets we consider. A significant
body of work is concerned with asynchronous or stop-and-go mixnets
that do not use forwarding rounds, but rely on random delays on
individual messages~\cite{diaz23provablysecure,loopix17mixnet}. A
different line of work relies on secure distributed computation
techniques like PIR or MPC to offer anonymous
communication~\cite{gibbs15riposte,chaum17cmix,kwon20xrd,Donghang23rpm,kwon17atom},
and the techniques they use are not directly comparable to the one we
presented here.

\section{Statement on the Use of Generative AI}
Generative AI tools, Claude, ChatGPT and Gemini, were used to revise
the text to eliminate typos, grammatical errors, to improve flow and
to make dedicated changes. Some simple but tedious algebraic
manipulation in proofs was also off-loaded to AI-agents. The Lean
formalization that accompanies this paper was carried out entirely by
Claude. The authors take responsibility for the entire work.

\bibliographystyle{plain}
\bibliography{upload_paper}

\appendix
\section{Proof of \cref{cor:indis_exit_msgs}}
\label{app:analysis-honest}

\begin{thm}\label{thm:inverse-binom-expec}
  Let $0 \leq p \leq 1$ be a real number, and $n > 1$ and $0 < k < n$
  be natural numbers. Let $X$ be a random variable distributed as
  $\mdistbinom{n-k}{p}$. Then,
  \[\expec{\frac{1}{X + k}} = \int_{0}^1 t^{k-1} (1-p+pt)^{n-k} \; dt\]
\end{thm}
\begin{proof}
  This is a standard result in probability theory. We present an
  elementary proof. First, observe that for any $a$,
  \[ \frac{1}{a} = \int_{0}^1 t^{a-1} \; dt \]
  In particular,
  \[ \frac{1}{X + k} = \int_{0}^1 t^{X+k-1} \; dt \]
  Taking expectations on both sides and commuting the expectation with
  the integral on the right using Fubini's theorem, we get:
  \begin{eqnarray*}
    \expec{\frac{1}{X + k}} & = & \int_{0}^1 \expec{t^{X+k-1}} \; dt \\
                            & = & \int_{0}^1 t^{k-1} \expec{t^X} \; dt 
  \end{eqnarray*}  
  Finally, observe that $t^X$ is the moment generating function of the
  binomial distribution $\mdistbinom{n-k}{p}$ and its expected value
  is known to be $(1-p+pt)^{n-k}$. Hence,
  \[ \expec{\frac{1}{X + k}} = \int_{0}^1 t^{k-1} (1-p+pt)^{n-k} \;
    dt\]
\end{proof}

\begin{cor}\label{thm:inverse-binom-expec-inst}
  \begin{enumerate}
  \item If $X$ is distributed as $\mdistbinom{n-1}{p}$, then
    \[ \expec{\frac{1}{X + 1}} = \frac{1-(1-p)^n}{np} \]
  \item If $X$ is distributed as $\mdistbinom{n-2}{p}$, then
    \[ \expec{\frac{1}{X + 2}} = \frac{np-1 + (1-p)^n}{n(n-1)p^2} \]
  \end{enumerate}
\end{cor}
\begin{proof} (1) Using \Cref{thm:inverse-binom-expec},
  \begin{eqnarray*}
    \expec{\frac{1}{X + 1}} & = & \int_{0}^1 t^{1-1} (1-p+pt)^{n-1} \;dt \\
                            & = & \int_{0}^1 (1-p+pt)^{n-1} \;dt \\
                            & = & \left.\frac{(1-p+pt)^{n}}{np}\right|_{t=0}^{t=1} \\
                            & = & \frac{1-(1-p)^n}{np}
  \end{eqnarray*}

  \medskip\noindent (2) Using \Cref{thm:inverse-binom-expec},
  \begin{eqnarray*}
    & & \expec{\frac{1}{X + 2}} \\
    & = & \int_{0}^1 t^{2-1} (1-p+pt)^{n-2} \;dt \\
    & = & \int_{0}^1 t (1-p+pt)^{n-2} \;dt \\
    & = & \left.\frac{t(1-p+pt)^{n-1}}{(n-1)p}\right|_{t=0}^{t=1} - \int_0^1 \frac{(1-p+pt)^{n-1}}{(n-1)p} \; dt \\
    & = & \frac{1}{(n-1)p} - \left. \frac{(1-p+pt)^{n}}{n(n-1)p^2} \right|_{t=0}^{t=1} \\
    & = & \frac{1}{(n-1)p} - \frac{1}{n(n-1)p^2} + \frac{(1-p)^n}{n(n-1)p^2}\\
    & = & \frac{np-1 + (1-p)^n}{n(n-1)p^2}
  \end{eqnarray*}

\end{proof}

By \cref{lem:odds-are-probs}, the conditional probabilities
compared in \cref{def:message-indistinguishability} are exactly the
posterior odds $\postpr{\ndepth}{\mind_1}$ and
$\postpr{\ndepth}{\mind_2}$. Unfolding the definition in these terms
gives the following restatement of \cref{cor:indis_exit_msgs}, which
bounds the adversary's privacy loss in both directions at once.

\begin{cor}[Restatement of \cref{cor:indis_exit_msgs}]\label{cor:indis_exit_msgs:re}
  The following restatement of \Cref{cor:indis_exit_msgs} holds: For
  any $\varepsilon > 0$ and any two distinct positions $\mind, \mindB$,
	\begin{equation*}
		\left| \log \left[\frac{\postpr{\ndepth}{\mind}}{\postpr{\ndepth}{\mindB}}\right] \right| \le \varepsilon
	\end{equation*}
	with probability at least $1 - \delta$ over $f \sim \partinf$,  for
	\begin{equation*}
		\delta = 2 (\nmsgs - 1) \cdot \beta^{\ndepth}\left[ 1 + \frac{\nmsgs}{2 \left(\nmsgs - 1\right) \sinh^2 \left(\frac{\varepsilon}{2}\right)}\right],
	\end{equation*}
\end{cor}
\begin{proof}
  First, we notice that the sum of all the $\postpr{\ndepth}{\mind}$s,
  i.e., $\sum_{\mind = 1}^\nmsgs \postpr{\ndepth}{\mind}$ is always
  equal to $1$. Hence,
  $\varnd{}{\sum_{\mind = 1}^\nmsgs \postpr{\ndepth}{\mind}} =
    0$. Using the formula for the variance of a sum of random
    variables,
    \[ \sum_{\mind = 1}^\nmsgs \varnd{}{\postpr{\ndepth}{\mind}} +
      \sum_{\mind, \mindB\in[1,\nmsgs] \conj \mind
          \neq \mindB} \covarn{\postpr{\ndepth}{\mind},
          \postpr{\ndepth}{\mindB}} = 0\]
      From the symmetry of the $\postpr{\ndepth}{\mind}$s, we get the
      covariance between $\postpr{\ndepth}{\mind}$ and
      $\postpr{\ndepth}{\mindB}$ for any $\mind, \mindB$:
	\begin{equation*}
		\covarn{ \postpr{\ndepth}{\mind}, \postpr{\ndepth}{\mindB} } = - \frac{\varnd{}{\postpr{\ndepth}{\mind}}}{\nmsgs - 1}.
	\end{equation*}
	Now, we consider
        \[\begin{array}{rcl}
          Y_1 & = & \postpr{\ndepth}{\mind} - e^{\varepsilon}\postpr{\ndepth}{\mindB} \\
          Y_2 & = & \postpr{\ndepth}{\mindB} - e^{\varepsilon}\postpr{\ndepth}{\mind}  
	\end{array}
        \]
	$Y_1$ and $Y_2$ are identically distributed random variables, each with mean $\frac{1}{N} \cdot \left(1 - e^{\varepsilon}\right)$ and variance
	\begin{equation*}
		\varnd{}{Y_i} = \varnd{}{\postpr{\ndepth}{\mind}} \cdot \left[ 1 + e^{2 \varepsilon}  \mathbin{+} \frac{2 e^{\varepsilon}}{\nmsgs - 1}\right]
	\end{equation*}
	We have to bound the probability of a privacy failure, i.e. the probability that $\postpr{\ndepth}{\mind} \ge e^{\varepsilon}\postpr{\ndepth}{\mindB}$ or $\postpr{\ndepth}{\mindB} \ge e^{\varepsilon}\postpr{\ndepth}{\mind}$, hence the probability that $Y_1 \ge 0 \disj Y_2 \ge 0$. Using the standard union bound, this probability is no more than $\delta = 2 \delta'$ where $\delta'$ is an upper bound on the probability of $Y_1 \ge 0$.

        Next, by Chebyshev's inequality,
	\begin{equation*}
		 \mathbb{P} \left(Y_1 \ge 0 \right)  \mathrel{\le} \frac{ \nmsgs^2 \varnd{}{Y_1}}{\left(e^{\varepsilon} - 1\right)^2} =: \delta'
	\end{equation*}
	hence
	\begin{align*}
          \delta' &= \nmsgs^2 \varnd{}{\postpr{\ndepth}{\mind}} \cdot \left[ \frac{\left(e^{\varepsilon} - 1\right)^2 + 2 e^{\varepsilon} + \frac{2e^{\varepsilon}}{\nmsgs - 1}}{\left(e^{\varepsilon} - 1\right)^2}\right] \\
		&= \nmsgs^2 \varnd{}{\postpr{\ndepth}{\mind}} \cdot \left[ 1 + \frac{2e^{\varepsilon} \frac{\nmsgs}{\nmsgs - 1}}{4e^{\varepsilon} \sinh^2 \left(\frac{\varepsilon}{2}\right)}\right] \\
		&= \nmsgs^2 \varnd{}{\postpr{\ndepth}{\mind}} \cdot \left[ 1 + \frac{\nmsgs}{2 \left( \nmsgs - 1\right) \sinh^2 \left(\frac{\varepsilon}{2}\right)}\right]
	\end{align*}
        Finally, the required $\delta$ is $2\delta'$:
	\begin{equation*}
		\delta = 2 \nmsgs^2 \varnd{}{\postpr{\ndepth}{\mind}} \cdot \left[ 1 + \frac{\nmsgs}{2 \left( \nmsgs - 1\right) \sinh^2 \left(\frac{\varepsilon}{2}\right)}\right]
	\end{equation*}
	From \Cref{thm:prob-var},  we know that the 
	variance of $\postpr{\ndepth}{\mind}$ is
	$\varnd{}{\postpr{\ndepth}{\mind}} = \beta^{\ndepth} \cdot
	\frac{1}{\nmsgs} \cdot \left( 1 - \frac{1}{\nmsgs}\right)$,
	respectively. Therefore, 
	\begin{equation*}
		\delta = 2 \beta^{\ell} \cdot \left(\nmsgs - 1\right) \cdot \left[ 1 + \frac{\nmsgs}{2 \left( \nmsgs - 1\right) \sinh^2 \left(\frac{\varepsilon}{2}\right)}\right],
	\end{equation*}
\end{proof}

\section{Further Details of the Differences from \texorpdfstring{\cite{kate25tightrope}}{Prior Work}}
\label{appendix:comparison-empirical}

In \cref{sec:evaluation}, we compared our provable bounds on $\dpde$
to the empirical estimates of prior work~\cite{kate25tightrope} only
for large values of $\dpep$, because the two works measure leakage
with different variants of differential privacy: our work uses
probabilistic differential privacy, while prior work uses approximate
differential privacy. This appendix explains the difference between
the two variants, and why it rules out a meaningful comparison for
small $\dpep$.

Both variants are stated in terms of the \emph{privacy loss} $L(o)$ of
an observation $o$: the logarithm of the ratio between the
probabilities of observing $o$ in the two worlds the adversary tries to
tell apart. In our setting, with the two worlds being that the message
of interest exits at position $\mind_1$ or at position $\mind_2$,
\cref{lem:odds-are-probs} and Bayes' rule give
$L(o) = \ln\bigl(\postpr{\ndepth}{\mind_1}/\postpr{\ndepth}{\mind_2}\bigr)$,
since every exit position is equally likely a priori. Our
\cref{def:message-indistinguishability} is a form of
\emph{probabilistic} differential privacy: $\dpde$ bounds the
probability that $L(o) > \dpep$, so every observation whose privacy
loss exceeds $\dpep$ counts fully towards $\dpde$. Prior work uses
\emph{approximate} differential privacy, whose $\dpde$ is the smallest
value such that $\Pr[o \in S \mid \mind_1] \leq e^{\dpep} \Pr[o \in S
\mid \mind_2] + \dpde$ for every set $S$ of observations. It equals
$\mathbb{E}\bigl[\max\bigl(0, 1 - e^{\dpep - L(o)}\bigr)\bigr]$, with
$o$ drawn from the first world, so an observation whose privacy loss
barely exceeds $\dpep$ contributes almost nothing. Since
$1 - e^{\dpep - L(o)} \leq 1$, the approximate $\dpde$ never exceeds
the probabilistic $\dpde$ at the same $\dpep$. Our provable bounds
therefore also bound the quantity that prior work estimates, but not
conversely.

How far apart the two variants are depends on $\dpep$, and this
determines where a comparison is meaningful. At $\dpep = 0$, they are
as far apart as possible. The approximate $\dpde$ is then the total
variation distance between the adversary's observations in the two
worlds, i.e., the expected gap between the two posteriors, which is
tiny when the posteriors differ only slightly. The probabilistic
$\dpde$, in contrast, is the probability that the two posteriors
differ at all, which is nearly $1$. Accordingly, our bound of
\cref{cor:indis_exit_msgs} grows without limit as $\dpep \to 0$, since
$\sinh^2(\dpep/2)$ appears in a denominator, and it would be
meaningless to compare it to the estimates of prior work for
$\dpep = 0$. As $\dpep$ grows, the two variants converge: at
$\dpep = \infty$, both equal the probability of a \emph{distinguishing
event}, an observation that rules out one of the two worlds entirely.
We therefore compare at large $\dpep$ only.

\end{document}